\documentclass[12pt]{article}

\pdfoutput=1

\usepackage{graphics}
\usepackage{epsfig}

\usepackage{amsfonts,amssymb,amsmath}
\usepackage{dsfont}
\usepackage{pifont}
\usepackage{multirow}
\usepackage{latexsym}
\usepackage{verbatim}
\usepackage{cite}
\usepackage{cancel}
\usepackage{slashed}
\usepackage{hyperref}
\usepackage{xcolor}
\usepackage{youngtab}

\usepackage{tikz}
\usetikzlibrary{shapes,arrows,positioning,snakes}

\usepackage[utf8]{inputenc}

\def\beq{\begin{equation}}
\def\eeq{\end{equation}}
\def\beqa{\begin{eqnarray}}
\def\eeqa{\end{eqnarray}}

\def\bfone{\relax{\rm 1\kern-.35em 1}}

\newcommand{\cR}{{\cal R}}

\newcommand{\cC}{{\cal C}}
\newcommand{\cM}{{\cal M}}
\newcommand{\cN}{{\cal N}}

\newcommand{\cH}{{\cal H}}
\newcommand{\cL}{{\cal L}}
\newcommand{\cV}{{\cal V}}

\newcommand{\bbR}{{\mathbb{R}}}

\newcommand{\bbZ}{{\mathbb{Z}}}

\newcommand{\SU}{{\text{SU}}}

\newcommand{\SL}{{\text{SL}}}
\newcommand{\SO}{{\text{SO}}}

\newcommand{\irrep}[2]{\mathbf{#1}_{(#2)}}

\newcommand{\be}{\begin{equation}}
\newcommand{\ee}{\end{equation}}
\newcommand{\ben}{\begin{displaymath}}
\newcommand{\een}{\end{displaymath}}
\newcommand{\bea}{\begin{eqnarray}}
\newcommand{\eea}{\end{eqnarray}}

\newcommand{\bean}{\begin{eqnarray*}}
\newcommand{\eean}{\end{eqnarray*}}

\newcommand{\D}{{\rm d}}

\DeclareMathAlphabet{\mathpzc}{OT1}{pzc}{m}{it}

\begin{document}
\pagestyle{plain}


\makeatletter \@addtoreset{equation}{section} \makeatother
\renewcommand{\thesection}{\arabic{section}}
\renewcommand{\theequation}{\thesection.\arabic{equation}}
\renewcommand{\thefootnote}{\arabic{footnote}}


\setcounter{page}{1} \setcounter{footnote}{0}


\begin{titlepage}
\begin{flushright}
\small YITP -- 19 -- 118
\end{flushright}

\bigskip

\begin{center}

\vskip 0cm

{\LARGE \bf 
6D (1,1) Gauged Supergravities\\[5pt]
from
\\[5pt]
Orientifold Compactifications 

}

\vskip 1.5cm

{\bf Giuseppe Dibitetto$^{1,2,3}$, Jose J. Fern\'andez-Melgarejo$^2$ \,and\, Masato Nozawa$^4$}
\\

\vskip 0.5cm

{\em 
$^1$ Departamento de F\'isica, Universidad de Oviedo,\\Avda.  Federico Garc\'ia Lorca s/n, 33007 Oviedo, Spain\\[1em]
$^2$ Departamento de F\'isica, Universidad de Murcia,\\Campus de Espinardo, E-30100 Murcia, Spain \\[1em]
$^3$ Institutionen f\"or fysik och astronomi, University of Uppsala,\\ Box 803, SE-751 08 Uppsala, Sweden \\[1em]
$^4$ Center for Gravitational Physics, Yukawa Institute for Theoretical Physics, \\ Kyoto University, Kyoto 606-8502, Japan}

\vskip 0.8cm

\end{center}

\vskip 1cm

\begin{center}

{\bf ABSTRACT}\\[3ex]

\begin{minipage}{13cm}
\small

We study dimensional reductions of M-theory/type II strings down to 6D in the presence of fluxes and spacetime filling branes and orientifold planes of different types. We classify all inequivalent orientifold projections giving rise to $\cN=(1,1)$ supergravities in 6D and work out the embedding tensor/fluxes dictionary for each of those. Finally we analyze the set of vacua for the different classes of reductions and find an abundance of ``no-scale'' type Minkowski vacua, as well as a few novel examples of (A)dS extrema.
\end{minipage}

\end{center}

\vfill

\end{titlepage}


\tableofcontents

\section{Introduction}
\label{sec:introduction}

The mechanism of flux compactifications appears to be essential in order to solve the issue of moduli stabiliazation within the context of dimensional reductions of string and M-theory. This procedure generically results in a lower dimensional effective supergravity theory with a non-trivial scalar potential inducing a mass for the excitations around a maximally symmetric vacuum, possibly with spontaneously broken supersymmetry.

Depending on the value of the effective cosmological constant ($\Lambda_{\mathrm{eff}}$), maximally symmetric vacua are divided into AdS ($\Lambda_{\mathrm{eff}}<0$), dS ($\Lambda_{\mathrm{eff}}>0$) or Mkw ($\Lambda_{\mathrm{eff}}=0$). While AdS vacua may be relevant in the context of the AdS/CFT correspondence, dS vacua describe accelerated cosmologies modeling dark energy and finally, Mkw vacua might provide candidate starting points for phenomenological constructions featuring supersymmetry breaking (\emph{e.g.} in the spirit of KKLT \cite{Kachru:2003aw}).    

Exploring the diversity of the string landscape in a top-down fashion is a problem of an enormous complexity \cite{Denef:2004ze,Denef:2006ad,Denef:2017cxt}, given the wide range of possible choices of geometrical and topological data of the internal manifolds. A crucial tool to explore large parts of this parameter space is given by consistent truncations, which allow us to trade this for the analysis of different lower dimensional effective descriptions in a bottom-up fashion instead.

While the consistency of truncations over a compact manifold generically requires a case-by-case study, we will mainly focus on a special class of manifolds which enjoy a group structure. In this particular setup, the consistency of the corresponding truncation automatically follows from group theoretical arguments. This construction is usually called twisted dimensional reduction \cite{Scherk:1979zr}.

Another crucial ingredient that will be considered in this work is spacetime filling orientifold planes. The inclusion of such extended objects with negative tension is argued to be required in order to evade the no-go theorem of \cite{Maldacena:2000mw} at a classical level and leave the possibility open to obtain a non-negative effective cosmological constant. With fluxes, internal geometry and sources at hand, the resulting lower dimensional description will be given by a gauged supergravity where the gauging is induced by the specific choice of background fluxes. Thanks to the recent developments in understanding and classifying all the possible consistent gauged supergravities facilitated by the advent of the so-called embedding tensor formalism \cite{Nicolai:2001sv,deWit:2002vt,deWit:2005ub}, a bottom-up approach provides extremely fruitful tools to investigate string vacua. An exhaustive classification of the gaugings of maximal supergravities for $D\ge8$ has been done \cite{Bergshoeff:2002nv,Bergshoeff:2003ri,FernandezMelgarejo:2011wx,deRoo:2011fa}, whereas for half-maximal theories, the analysis extends to $D\ge7$ \cite{Dibitetto:2012rk,Dibitetto:2015bia}.

Gaugings and massive deformations are the unique prescriptions for the deformation of extended supergravity theories (some comprehensive reviews are found in \cite{Samtleben:2008pe,Trigiante:2016mnt}). 
The resulting gauged supergravities admit non-Abelian gauge groups, fermion mass terms, as well as a scalar potential. 
The embedding tensor specifies how the gauge group is embedded into the duality group,  and allows us to construct all the possible gauged supergravity theories in a duality covariant fashion. The embedding tensor should satisfy the linear and quadratic constraints: the former is required by supersymmetry and the latter comes from the consistency of the deformation.

In the light of the aforementioned connection between flux backgrounds and gaugings, one may then be tempted to hope that all the lower dimensional supergravities can be obtained from a suitable compactification of string/M-theory. Unfortunately, as of now this still remains an open question.
However, various implementations of a duality covariant formalism in string theory naturally seem to go beyond geometry in a strict sense. Along these lines the so-called non-geometric fluxes were originally introduced in~\cite{Shelton:2005cf}.


Returning to the case of lower dimensional theories with a known higher dimensional origin, a first substantial progress in understanding the embedding tensor/fluxes dictionary was made in \cite{Roest:2009dq,DallAgata:2009wsi,Dibitetto:2010rg,Dibitetto:2011gm} within the context of $D=4$ $\mathcal{N}=4$ supergravities (in the formulation of \cite{Schon:2006kz}) arising from orientifold reductions of type II strings on a twisted $T^6$ with fluxes. Subsequently, in \cite{Dibitetto:2012ia,Dibitetto:2014sfa}, the extra conditions obstructing an embedding within $\mathcal{N}=8$ supergravity were identified with tadpoles for spacetime filling BPS sources. Though extremely valuable at a conceptual level, the above treatment in four dimensions does permit a systematic exploration of the set of string vacua. This is due to the large number of flux components, which cause a dramatic increase of the complexity of the problem at hand. The actual exhaustive vacua scan was only possible within a particular sector of the theory enjoying SO(3) invariance.

Motivated by this, we will now focus on the very same issue but in the context of half-maximal supergravities in six dimensions, where we expect far smaller amounts of flux parameters, due to the presence of smaller global symmetries. This particular setup will first of all, allow us to classify all inequivalent orientifold projections which are consistent with $(1+5)$D Lorentz symmetry and within perturbative control. This will yield a subset of what was found in the classification of \cite{Dibitetto:2018wvc}, where also exotic objects were considered. Furthermore it will allow for a systematic treatment of the vacua scan. 

With this minimal set of compactification ingredients and the embedding tensor techniques as a toolbox, it is technically possible to exhaustively explore this portion of the string landscape and find new interesting examples of Mkw, (non-)supersymmetric AdS as well as dS extrema. These will serve as possible tests for our current understanding of a consistent quantum gravity theory and its rules. One could \emph{e.g.} test nonperturbative instabilities of non-supersymmetric vacua as envisioned by \cite{Ooguri:2016pdq,Danielsson:2016mtx}, or question the (non-)existence of dS vacua as discussed in \cite{Obied:2018sgi,Danielsson:2018ztv}, both at a perturbative and nonperturbative level.

In this paper, we discuss flux compactifications of string/M-theory down to six dimensions with localized sources that explicitly break half the supersymmetry. This includes various different orientifolds in (massive) Type IIA, as well as Type IIB and M-theory. The range of inequivalent possibilities is summarized in Figure \ref{fig:branes-chart}. We then give an encyclopedic relation between flux elements and embedding tensor components for individual cases. Considering the configurations of embedding tensor corresponding to the given compactifications, we attempt to systematically find critical points of the scalar potential. In most cases, our analysis turns out to be exhaustive. We note that in \cite{Passias:2012vp,Apruzzi:2014qva}, the existence of six-dimensional AdS solutions preserving some supersymmetry has been studied. 
\tikzstyle{theory} = [draw, text width=5em, rounded corners, text centered, node distance=4cm, minimum height=3em, fill=red!20]
\tikzstyle{theory6} = [draw, text width=5em, rounded corners, text centered, node distance=4cm, minimum height=3em, fill=green!20]
\tikzstyle{orientifold} = [ellipse, draw, fill=blue!20, 
    text width=4em, text centered, minimum height=3em,node distance=4cm]
\tikzstyle{line} = [draw, -latex']
   
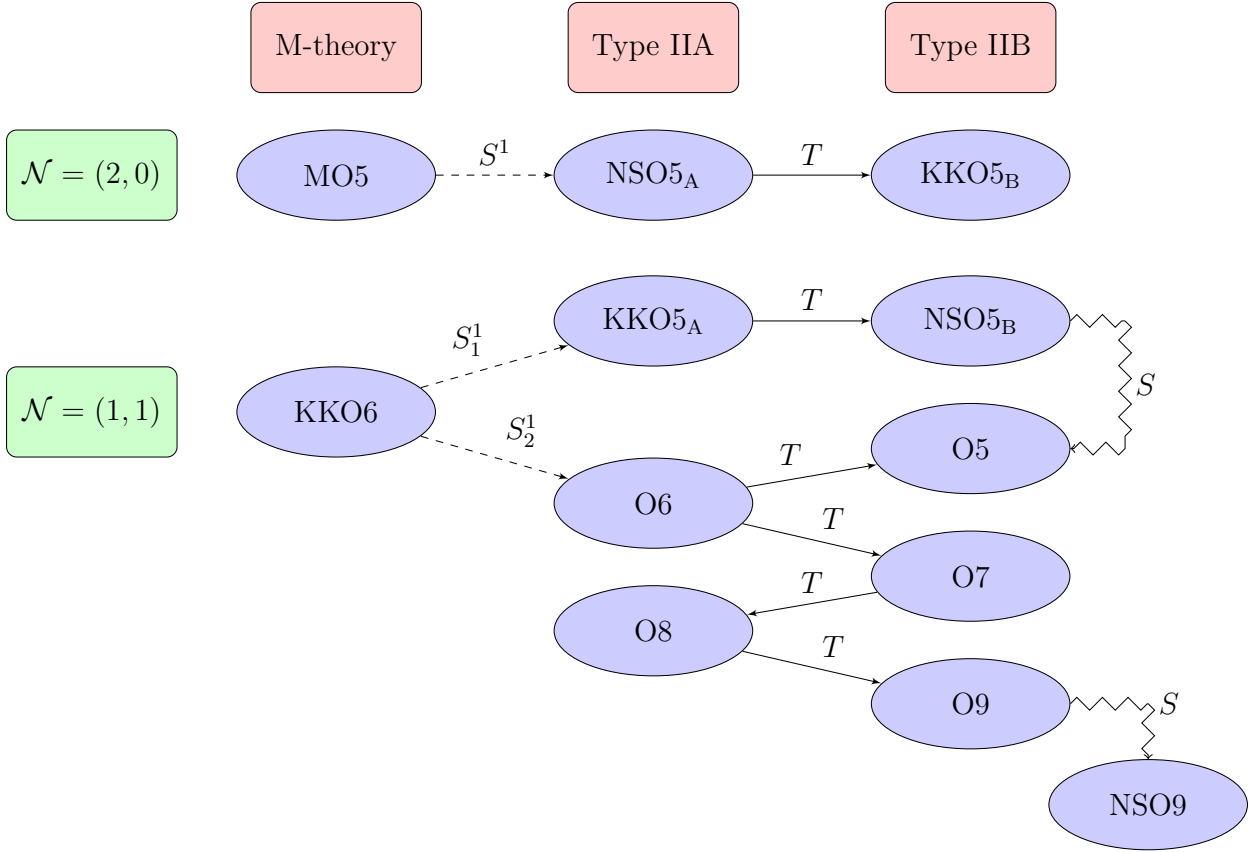
\begin{figure}[t!]
\centering
\resizebox{\textwidth}{!}{%
\begin{tikzpicture}[node distance = 4cm, auto]
	\node [theory] (mtheory) {M-theory};
	\node [theory,right = 2 cm of mtheory] (IIA) {Type IIA};
	\node [theory,right = 2 cm of IIA] (IIB) {Type IIB};
	\node [theory6, below left = .5 cm and 1 cm  of mtheory] (iib) {$\cN=(2,0)$};
	\node [orientifold,below = .5 cm of mtheory] (MO5) {MO5};
	\node [orientifold,below = .5 cm of IIA] (NSO5A) {NSO5$_\text{A}$};
	\node [orientifold,below = .5 cm of IIB] (KKO5B) {KKO5$_\text{B}$};
	\node [orientifold,below = .75 cm of NSO5A] (KKO5A) {KKO5$_\text{A}$};
	\node [orientifold,below = .75 cm of KKO5B] (NSO5B) {NSO5$_\text{B}$};
	\node [theory6, below = 2 cm of iib] (iia) {$\cN=(1,1)$};
	\node [orientifold, below = .5 cm of NSO5B] (O5) {O5};
	\node [orientifold, below =  1.25 cm of KKO5A] (O6) {O6};
	\node [orientifold, below =  .5 cm of O5] (O7) {O7};
	\node [orientifold, below = .5 cm of O6] (O8) {O8};
	\node [orientifold, below = .5 cm of O7] (O9) {O9};
	\node [orientifold, below = 2 cm of MO5] (KKO6) {KKO6};
	\node [orientifold, below right = .5 cm and .5 cm of O9] (NSO9) {NSO9};
	\path [line,dashed] (MO5) -- node {$S^1$} (NSO5A);
	\path [line] (NSO5A) -- node {$T$} (KKO5B);
	\path [line,dashed] (KKO6) -- node {$S^1_1$} (KKO5A);
	\path [line,dashed] (KKO6) -- node {$S^1_2$} (O6);
	\path [line] (KKO5A) -- node {$T$} (NSO5B);
	\path [line] (O6) -- node {$T$} (O5);
	\path [line] (O6) -- node {$T$} (O7);
	\coordinate[right = .75 cm of NSO5B] (arrowS1);
	\draw [snake=zigzag] (NSO5B) -- (arrowS1) ;
	\draw [->,snake=zigzag] (arrowS1) |- node [near start]{$S$} (O5);
	\path [line] (O7) -- node [above] {$T$} (O8);
	\path [line] (O8) -- node {$T$} (O9);
	\draw [->,snake=zigzag] (O9) -| node {$S$} (NSO9);
\end{tikzpicture}}
\caption{\textit{
The various type II/M-theory orientifold compactifications that give rise to either $\cN=(2,0)$ or $\cN=(1,1)$ supergravity theories and the relation among them: $S^1$ stands for a compactification on a circle, whereas $T$ and $S$ refer to T- and S-duality, respectively. The NSO9 plane is a solution of heterotic supergravity and has not been discussed in this work.
}}
\label{fig:branes-chart}
\end{figure}

This work is organized as follows. In Section \ref{sec:compactification} we derive the scalar potential arising from type II and M-theory reductions to six dimensions in the presence of local sources. In Section \ref{sec:gauged-sugra} we introduce $\cN=(1,1)$ $D=6$ supergravity and its consistent deformations by following the embedding tensor formalism. Section \ref{sec:orientifold} contains the orientifold compactifications of type II and M-theory shown in Figure \ref{fig:branes-chart} and a discussion of the critical points for each case. Finally, in Section \ref{sec:conclusions} we present our conclusions and discuss other further developments. Some technical auxiliary material is collected in Appendices \ref{app:truncation} \& \ref{app:global}.

\section{Deriving the scalar potential}
\label{sec:compactification}

In this section we will derive the scalar potential arising from the compactification of the bosonic sector of type II and M-theory in the presence of local sources. In particular, we will calculate the contribution of each term in the 10/11-dimensional action to an effective moduli potential and calculate the functional dependence of the universal moduli.

\subsection{Reductions of Type II down to 6D}
\label{sec:II-compactification}

Let us consider the bosonic part of the action of type II supergravities in the string frame\footnote{We retain conventions where $2\kappa_{10}^2\,=\,1$.}
\begin{align}
S^{\text{II}}
=
\int \D^{10}x\sqrt{-g^{(10)}}\left(
	e^{-2\Phi}\left(
		\mathcal{R}^{(10)}
		+4(\partial\Phi)^2
		-\frac1{12}|H_{(3)}|^2
		\right)
	- \sum_p\frac{1}{2p!}|F_{(p)}|^2
	\right) + S_{\mathrm{CS}}
\ ,
\label{eq:II-action}
\end{align}
where $p=0,2,4$ for massive type IIA and $p=1,3,5$ for type IIB theory, while $S_{\mathrm{CS}}$ denotes a topological term whose explicit form is different in the IIA/IIB cases. $|F_{(p)}|^2$ denotes contraction of all indices with respect to the 10-dimensional metric. In addition, we add local sources such as spacetime filling D$p$-branes and O$p$-planes, which contribute to the action via the term
\begin{align}
S^{(\text{O}p/\text{D}p)}
=
-T_p\int_{\cC_{p+1}} \D^{p+1}x\sqrt{-\tilde g^{(p+1)}} e^{-\Phi} 
\ ,
\label{eq:source-term}
\end{align}
where $T_p$ represents the tension of the corresponding extended object, ${\cC_{p+1}}$ its worldvolume and $\tilde g^{(p+1)}$ is the pull-back of the 10-dimensional metric on the worldvolume.

To perform the dimensional reduction down to $D=6$, we need to introduce a parameterization of the metric $g^{(10)}$ in terms of the 6-dimensional non-compact metric and the moduli describing deformations of the 4-dimensional internal metric. By choosing
\begin{align}
\D s_{(10)}^2
=
g_{MN}^{(10)}\D x^M\otimes \D x^N
=
\tau^{-2} g_{\mu\nu}^{(6)}\D x^\mu\otimes \D x^\nu
+\rho \, \D s_4^2
\ ,
\label{eq:II-metric}
\end{align}
where the internal metric $g^{(4)}$ is normalized such that $\int \D ^4y \sqrt{g^{(4)}}=1$, the so-called universal moduli $\rho$ and $\tau$ are singled out, whereas the other moduli are still sitting inside $g^{(4)}$ and describe volume preserving deformations of the internal geometry. 
We introduce local flat indices $m,n$ as 
\begin{align}
\D s_4^2
=
\cM_{mn} e^m\otimes e^n 
\ ,
\end{align}
where the matrix $\cM_{mn}$ parameterizes the coset SL(4$,\bbR$)/SO(4) and in particular $\det\cM=1$.

The requirement of obtaining the $D=6$ gravity action in the Einstein frame after the compactification procedure implies the following constraint \cite{Hertzberg:2007wc}
\begin{align}
\rho^2 
\stackrel{!}{=}
e^{2\Phi}\tau^4
\ .
\label{eq:II-Einstein-condition}
\end{align}
Therefore, the universal moduli $(\rho,\tau)$ fix the internal volume as well as the string coupling.


Let us now consider the dependence of the various contributions to the scalar potential on $(\rho,\tau)$ based on the parameterization we have introduced in \eqref{eq:II-metric}. The 10-dimensional Ricci scalar reduces to the following leading part (\emph{i.e.}, up to terms involving derivatives of the moduli)
\begin{align}
\cR^{(10)} 
\quad
\longrightarrow
\quad
\tau^2 \cR^{(6)}
+\rho^{-1} \cR^{(4)}
\ ,
\end{align}
whereas the determinant of the metric reduces to
\begin{align}
\sqrt{-g^{(10)}}
\quad
\longrightarrow
\quad
\tau^{-6} \rho^2 \sqrt{g^{(4)}} \sqrt{-g^{(6)}}
\ .
\end{align}

First of all, the reduction of the Einstein term inside \eqref{eq:II-action} will give rise to the gravity action in six dimensions in the Einstein frame plus a first contribution to the scalar potential which we denote by $V_\omega$, where $\omega$ represents the \emph{metric flux}. Calculating this explicitly, we have
\begin{eqnarray}
\int \D ^{10}x \sqrt{-g^{(10)}} e^{-2\Phi} \cR^{(10)} 
\ \ &
\longrightarrow
&\ \ 
\int \D ^6 x \sqrt{-g^{(6)}}(
	\underbrace{\tau^{-4}\rho^2 e^{-2\Phi}}_{=1\text{ see \eqref{eq:II-Einstein-condition}}} \ \cR^{(6)}
	+\underbrace{\tau^{-6}\rho e^{-2\Phi}}_{=\rho^{-1}\tau^{-2}} \ \cR^{(4)}
	)
\nonumber\\
&=&\ \  
\int \D ^6 x \sqrt{-g^{(6)}}\big(
	\cR^{(6)}
	+\underbrace{\rho^{-1} \tau^{-2} \ \cR^{(4)}}_{-V_\omega}
	\big)
\ ,
\end{eqnarray}
where $V_\omega\equiv-\rho^{-1} \tau^{-2} \ \cR^{(4)}$.
The expression of $\cR^{(4)}$ in twisted toroidal compactifications can be written as \cite{Scherk:1979zr}
\begin{align}
\cR^{(4)}
=
-\frac14 \cM_{mq} \cM^{nr} \cM^{ps} \omega_{np}{}^q \omega_{rs}{}^m
-\frac12 \cM^{np}\omega_{mn}{}^q \omega_{qp}{}^m
\ ,
\end{align}
where the matrix $\cM^{mn}$ denotes the inverse of $\cM_{mn}$ and the components $\omega_{mn}{}^p$ represent the structure constants of the corresponding group manifold chosen for the compactification. As such, they must satisfy a unimodularity constraint as well as the Jacobi identities for closure of the underlying Lie algebra (see  Appendix \ref{app:global} for details)
\begin{align}
\omega_{mn}{}^n=0 \ , \qquad \mathrm{and} \qquad \omega_{[mn}{}^r\omega_{p]r}{}^q
=
0
\ .
\label{eq:II-Jacobi}
\end{align}

A further contribution to the scalar potential comes from the $H$ flux; reducing the corresponding term in the action \eqref{eq:II-action} yields
\begin{align}
\int \D^{10}x \sqrt{-g^{(10)}} \left(
	-\frac{1}{12} e^{-2\Phi}|H_{(3)}|^2
	\right)
\quad
\longrightarrow
\quad
\int \D^6 x \sqrt{-g^{(6)}}\underbrace{ \left(
	-\frac{1}{12} H_{mnp} H^{mnp} \rho^{-3} \tau^{-2}
	\right)}_{-V_H}
\ ,
\end{align}
where $V_H\equiv \frac{1}{12} H_{mnp} H^{mnp} \rho^{-3} \tau^{-2}$ and the contraction on the indices $m$, $n$, $p$ is intended to be w.r.t. the internal metric $g^{(4)}$.

The R-R $p$-forms contribute to the scalar potential as follows
\begin{align}
\int \D^{10}x \sqrt{-g^{(10)}} \left(
	-\frac{1}{2p!} |F_{(p)}|^2
	\right)
\quad
\longrightarrow
\quad
\int \D^6x \sqrt{-g^{(6)}}\underbrace{\left(
	-\frac{1}{2p!} F_{m_1\ldots m_p}F^{m_1\ldots m_p} \rho^{2-p}\tau^{-6}
	\right)}_{-V_{F_p}}
\ ,
\end{align}
where $V_{F_p}\equiv \frac{1}{2p!} F_{m_1\ldots m_p}F^{m_1\ldots m_p} \rho^{2-p}\tau^{-6}$.

Finally a last contribution to the scalar potential arises from the reduction of the local source term in the 10-dimensional action given in \eqref{eq:source-term}. Such a reduction yields\footnote{Please note that the O-plane and the corresponding D-branes should wrap the non-compact 6-dimensional space completely (this implies $p \ge 5$) in order for 1+5 dimensional Lorentz symmetry to be preserved.}
\begin{align}
-T_p\int_{\cC_{p+1}} \D^{p+1}x\sqrt{-\tilde g^{(p+1)}} e^{-\Phi}
\quad
\longrightarrow
\quad
\int \D^6x \sqrt{-g^{(6)}}\underbrace{\left(
	-T_p \rho^{\frac{p-7}{2}}\tau^{-4}\text{vol}_{p-5}
	\right)}_{-V_{\text{O}p/\text{D}p}}
\ ,
\end{align}
where $\text{vol}_{p-5}\equiv\int_{\tilde \cC_{p-5}} \D^{p-5}y\sqrt{\tilde g^{(p-5)}}$ defines the interval volume wrapped by the O$p/$D$p$ system and the contribution to the potential is $V_{\text{O}p/\text{D}p}\equiv T_p \rho^{\frac{p-7}{2}}\tau^{-4}\text{vol}_{p-5}$.

Since moreover no extra contributions comes from the 10D topological term, the reduced $D=6$ theory is described by the following effective Lagrangian 
\begin{align}
\label{D6Lag}
{\cal L}_6 = \sqrt{-g^{(6)}}\,\left(\cR^{(6)} +2{\cal L}_{\rm kin}-V\right) \,, 
\end{align}
where $V$ denotes the full scalar potential 
\begin{align}
V=V_H+V_\omega+\sum_p V_{F_p}+V_{\text{O}p/\text{D}p}
\ .
\label{eq:II-potential}
\end{align}
The scalar fields span a $\mathbb R^+_{\rho} \times \mathbb R^+_{\tau} \times {\rm SL}(4,\bbR)/{\rm SO}(4)$ geometry
and the corresponding kinetic Lagrangian reads 
\begin{align}
\label{}
{\cal L}_{\rm kin} =-\frac{(\partial\rho)^2}{2\rho^2}-\frac{2(\partial\tau)^2}{\tau^2} +\frac 18 {\rm Tr}(\partial {\cal M}\partial {\cal M}^{-1}) \ . 
\end{align}

\subsection{Reductions of M-theory down to 6D}
\label{sec:11d-compactification}

Let us now analyze the bosonic action of 11-dimensional supergravity
\begin{align}
S^{\text{11}}
=
\int \D^{11}x\sqrt{-g^{(11)}}\left(
	\cR^{(11)}
	-\frac{1}{2\cdot 4!}|G_{(4)}|^2
	\right) 
\ ,
\label{eq:11d-action}
\end{align}
where $|G_{(4)}|^2$ denotes contraction of all indices with respect to the 11-dimensional metric. 
The only spacetime filling sources that we will be considering in this case are KK monopoles. Since these objects are directly sourced by the metric, their contribution already comes through the 11D Einstein-Hilbert term and no extra source terms are needed in the 11D action.

To perform the dimensional reduction down to $D=6$, we need to introduce a parameterization of the metric $g^{(11)}$ in terms of the 6-dimensional non-compact metric and the moduli describing the 5-dimensional internal metric. We choose
\begin{align}
\D s_{(11)}^2
=
g_{MN}^{(11)}\D x^M\otimes \D x^N
=
\tau^{-2} g_{\mu\nu}^{(6)}\D x^\mu\otimes \D x^\nu
+\rho \, \D s_5^2
\ ,
\label{eq:11d-metric}
\end{align}
where the internal metric $g^{(5)}$ is normalized such that $\int \D ^5y \sqrt{g^{(5)}}=1$, whereas other volume preserving moduli are still sitting inside $g^{(5)}$. 
We introduce 5-dimensional local flat indices $\hat m,\hat n$ as 
\begin{align}
\D s_5^2
=
\hat\cM_{mn} \hat e^{\hat m}\otimes \hat e^{\hat n}
\ ,
\end{align}
where the matrix $\hat \cM_{\hat m\hat  n}$ parameterizes the coset SL(5$,\bbR$)/SO(5) and in particular $\det\hat\cM=1$.
The requirement of having the $D=6$ gravity action in the Einstein frame after the compactification procedure implies
\begin{align}
\tau^{-4}\rho^{5/2}
\stackrel{!}{=}
1
\ ,
\label{eq:11d-Einstein-condition}
\end{align}
which reduces the set of universal moduli to the only $\rho$ which has the role of fixing the internal volume.


Let us now consider the dependence of the various contributions to the scalar potential on $\rho$ based on the parameterization we have introduced in \eqref{eq:II-metric}. The 11-dimensional Ricci scalar reduces to the following leading part (\emph{i.e.}, up to terms involving derivatives of the moduli)
\begin{align}
\cR^{(11)} 
\quad
\longrightarrow
\quad
\tau^2 \cR^{(6)}
+\rho^{-1} \cR^{(5)}
\ ,
\end{align}
whereas the determinant of the metric reduces to
\begin{align}
\sqrt{-g^{(11)}}
\quad
\longrightarrow
\quad
\tau^{-6} \rho^{5/2} \sqrt{-g^{(6)}}
\ .
\end{align}

First of all, the reduction of the Einstein term inside \eqref{eq:11d-action} will give rise to the gravity action in six dimensions in the Einstein frame plus a first contribution to the scalar potential $V_\omega$, associated to the \emph{metric flux}. Calculating this explicitly, we have
\begin{eqnarray}
\int \D ^{11}x \sqrt{-g^{(11)}}  \cR^{(11)} 
\ \ &
\longrightarrow
&\ \ 
\int \D ^6 x \sqrt{-g^{(6)}}(
	\underbrace{\tau^{-4}\rho^{5/2} }_{=1\text{ see \eqref{eq:11d-Einstein-condition}}} \ \cR^{(6)}
	+\tau^{-6}\rho^{3/2}  \ \cR^{(5)}
	)
\nonumber\\
&=&\ \  
\int \D ^6 x \sqrt{-g^{(6)}}(
	\cR^{(6)}
	+\underbrace{\tau^{-6}\rho^{3/2} \ \cR^{(5)}}_{-V_\omega}
	)
\ ,
\end{eqnarray}
where $V_\omega\equiv -\rho^{-6} \ \cR^{(5)}$, by virtue of \eqref{eq:11d-Einstein-condition}.
The expression of $\cR^{(5)}$ in twisted toroidal compactifications can be written as 
(assuming unimodularity of the group)
\begin{align}
\cR^{(5)}
=
-\frac14 \hat\cM_{\hat m\hat q} \hat \cM^{\hat n\hat r} \hat \cM^{\hat p\hat s} \omega_{\hat n\hat p}{}^{\hat q} \omega_{\hat r\hat s}{}^{\hat m}
-\frac12 \hat\cM^{\hat n\hat p}\omega_{\hat m\hat n}{}^{\hat q} \omega_{\hat q\hat p}{}^{\hat m}
\ ,
\end{align}
where the matrix $\hat \cM^{\hat m\hat n}$ denotes the inverse of $\hat \cM_{\hat m\hat n}$ and $\omega_{\hat m\hat n}{}^{\hat p}$ represents the structure constants of the corresponding group manifold chosen for the compactification, which are therefore still subject to the Jacobi identities.

The 3-form contributes to the scalar potential as follows 
\begin{align}
\int \D^{11}x \sqrt{-g^{(11)}} \left(
	-\frac{1}{2\cdot 4!} |G_{(4)}|^2
	\right)
\quad
\longrightarrow
\quad
\int \D^6x \sqrt{-g^{(6)}}\underbrace{\left(
	-\frac{1}{2\cdot4!} G_{\hat m_1\hat m_2\hat m_3\hat m_4}G^{\hat m_1\hat m_2\hat m_3\hat m_4} \rho^{{-3/2}}\tau^{-6}
	\right)}_{-V_{G_4}}
\ ,
\end{align}
where $V_{G_4}\equiv \frac{1}{2\cdot4!} G_{\hat m_1\hat m_2\hat m_3\hat m_4}G^{\hat m_1\hat m_2\hat m_3\hat m_4}{\rho^{{-21/4}}}$, upon using \eqref{eq:11d-Einstein-condition}.


In summary, the reduced six dimensional Lagrangian takes the form (\ref{D6Lag}), 
where the kinetic term for the $\mathbb R^+_\rho \times {\rm SL}(5,\bbR)/{\rm SO}(5) $ scalars is now parameterized as
\begin{align}
\label{}
{\cal L}_{\rm kin}=-\frac{45}{32}\frac{(\partial \rho)^2}{\rho^2}+\frac 18 {\rm Tr}(\partial \hat{\cal M} \partial \hat{\cal M}^{-1})\ . 
\end{align}
The potential will be given by
\begin{align}
V= V_\omega + V_{G_4}
\ .
\label{eq:11d-potential}
\end{align}

We will establish a mapping between flux compactifications of type II and M-theory with O$p$/D$p$-branes for $p\ge5$ and half-maximal 6-dimensional gauged supergravities. Depending on the type of orientifold projection considered, the obtained theory will be either iia ($\cN=(1,1)$, \emph{i.e.} nonchiral) or iib ($\cN=(2,0)$, \emph{i.e.} chiral). In the diagram of Figure \ref{fig:branes-chart} we summarize the various compactifications with sources that can be performed, their relations through string dualities and the supergravity theories that they give rise to.

\section{Gauged supergravity formulation}
\label{sec:gauged-sugra}

In this section we would like to interpret the orientifold compactifications mentioned above as supergravity theories in six dimensions subject to embedding tensor deformations. Since chiral supergravities do not allow for any such deformations, the spacetime filling orientifold planes that we consider are those that truncate type II/M-theory to a half-maximal iia supergravity in $D=6$ (\emph{i.e.,} $\cN=(1,1)$). This will allow us to match the scalar potentials derived in \eqref{eq:II-potential} and \eqref{eq:11d-potential} with a supergravity potential induced by a certain gauging which involves all the scalars beyond those sitting in the metric.


Type iia half-maximal supergravities in $D=6$ enjoy $G=\mathbb{R}^+\times\text{SO}(4,4)$ global symmetry \cite{Bergshoeff:2007vb}. General global symmetry transformations inside $G$ include  global $\mathbb{R}^+$ rescalings as well as T-duality transformations. The scalar fields span the coset\footnote{We will denote by $M,N,\cdots$ fundamental SO$(4,4)$ indices, which are raised and lowered by the SO$(4,4)$ metric in light-cone coordinates $\eta_{MN}\equiv\begin{pmatrix}
0 &\mathds{1}_4\\ \mathds{1}_4 & 0
\end{pmatrix}$.}
\begin{align}
\underbrace{\mathbb{R^+}}_{\Sigma}\ \times \ \underbrace{\frac{\text{SO}(4,4)}{\text{SO}(4)\times \text{SO}(4)}}_{\cH_{MN}}
\ ,
\label{eq:sugra-coset}
\end{align}
where $\Sigma$ has charge $-1$, whereas the scalar matrix $\cH_{MN}$ is neutral w.r.t. the aforementioned rescalings. Let us introduce the vielbein $\cV_M{}^{\underline{M}}$ such that 
\begin{align}
\cV_M{}^{\underline{M}}\cV_N{}^{\underline{M}}
\equiv
\cV_M{}^{\underline{m}}\cV_N{}^{\underline{m}}
+\cV_M{}^{\underline{\hat m}}\cV_N{}^{\underline{\hat m}}
=
\cH_{MN}
\ ,
\end{align}
where $\underline{M}=(\underline{m},\underline{\hat m})$ denotes a local $\text{SO}(4)\times \text{SO}(4)$ index and splits into its timelike and spacelike parts respectively. The kinetic Lagrangian is given by
\begin{align}
\cL_{\text{kin}}
=
-2\Sigma^{-2}(\partial\Sigma)^2
+\frac{1}{16}\partial\cH_{MN}\partial\cH^{MN}
\ .
\end{align}

The consistent deformations of the theory can be encoded in the so-called embedding tensor
\begin{align}
\Theta
=
\underbrace{{\bf 8}_c^{(+3)}}_{p=2}
\ \oplus \
\underbrace{{\bf 8}_c^{(-1)} \ \oplus \ {\bf 56}_c^{(-1)}}_{p=1}
\ ,
\end{align}
which comprises a massive deformation ($p=2$ type) as well as some gaugings ($p=1$ type) in the $\mathbb{R}^+$ and SO$(4,4)$ part, respectively \cite{Bergshoeff:2007vb}.
To describe the different embedding tensor irrep's, 
let us introduce the following notation
\begin{align}
\zeta_M\in {\bf 8}_c^{(+3)}
\ , \quad
\xi_M\in {\bf 8}_c^{(-1)}
\ , \quad
f_{[MNP]}\in {\bf 56}_c^{(-1)}
\ ,
\end{align}
where $f_{[MNP]}$ plays the role of generalized structure constants. 

The closure of the gauge algebra and the consistency of the massive deformation imply a set of quadratic constraints (QC) on the embedding tensor which are given by
\begin{align}
\begin{array}{rccrcc}
3f_{R[MN}f_{PQ]}{}^R - 2f_{[MNP}\xi_{Q]}
\ =\  0
&
(\mathbf{35}_{v}^{(-2)}\oplus\mathbf{35}_{s}^{(-2)})
&,\quad
&
\zeta_{(M}\xi_{N)}
\ =\  0
&
(\mathbf{35}_{c}^{(+2)}\oplus\mathbf{1}^{(-2)})
&,\quad 
\\[5pt]
f_{MNP}\zeta^P-\xi_{[M}\zeta_{N]}
\ =\  0
&
(\mathbf{28}^{(+2)})
&,\quad
&
\xi_M\xi^M
\ =\  0
&
(\mathbf{1}^{(-2)})
&,\quad
\\[5pt]
f_{MNP}\xi^P
\ =\  0
&
(\mathbf{28}^{(-2)})
&,\quad
&
\zeta_M\xi^M
\ =\  0
&
(\mathbf{1}^{(+2)})
&.\quad 
\end{array}
\label{eq:QC}
\end{align}

One important consequence of the gauging procedure is that it induces the following scalar potential
(see also \cite{Dibitetto:2018iar})
\begin{multline}
V 
=
\tfrac{g^2}{4}\left[
	f_{MNP}f_{QRS}\Sigma^{-2}\left(
		\tfrac{1}{12}\cH^{MQ}\cH^{NR}\cH^{PS}
		-\tfrac14 \cH^{MQ}\eta^{NR}\eta^{PS}
		+\tfrac16 \eta^{MQ}\eta^{NR}\eta^{PS}
		\right)
	\right.
	\\
	\left.
	+\tfrac12 \zeta_M\zeta_N \Sigma^6\cH^{MN}
	+\tfrac23 f_{MNP}\zeta_Q\Sigma^2 \cH^{MNPQ}
	+\tfrac54 \xi_M\xi_N \cH^{MN} \Sigma^{-2}
	\right]
\ ,
\label{eq:sugra-potential}
\end{multline}
where $\cH^{MN}$ denotes the inverse of $\cH_{MN}$ and $\cH_{MNPQ}\equiv\epsilon_{\underline{mnpq}}\cV_M{}^{\underline{m}}\cV_N{}^{\underline{n}}\cV_P{}^{\underline{p}}\cV_Q{}^{\underline{q}}$. The above scalar potential can be obtained as a $\mathbb{Z}_2$ truncation of the maximal theory in six dimensions, \emph{i.e.}, $\cN=(2,2)$ \cite{Bergshoeff:2007ef} and compared to that one of the half-maximal theory in $D=5$ \cite{Schon:2006kz} upon a reduction on a circle $S^1$. In particular, in order for an $\cN=(1,1)$ gauging to admit an embedding within the maximal theory, it needs to satisfy the following two extra QC
\begin{align}
\left\{\begin{array}{rccc}
f_{MNP}f^{MNP} &=&0&,
\\[5pt]
\left.f_{[MNP}\zeta_{Q]}\right|_{\text{SD}} &=&0&,
\end{array}
\right.
\label{eq:QC-extra}
\end{align}
where $|_{\text{SD}}$ denotes the self-dual part of a four-form, in analogy with the $D=4$ case (see \cite{Dibitetto:2011eu,Dibitetto:2012ia}).
We defer the detailed derivation for (\ref{eq:QC}) and (\ref{eq:QC-extra}) to Appendix \ref{app:truncation}.

In what follows, we will be extremizing the scalar potential \eqref{eq:sugra-potential} specified for gaugings which are interpreted as coming from certain orientifold reductions. Once in an extremum $\phi_0$ of $V$, one needs to discuss its physical properties, such as \emph{e.g.} its mass spectrum. 
To this end, we use the following formula 
\begin{align}
\label{}
(m^2)^\alpha{}_\beta =2 K^{\alpha\gamma}\partial_\beta \partial_\gamma V|_{\phi_0} \,, \qquad 
{\cal L}_{\rm kin}= -\frac 12 K_{\alpha \beta}\partial \phi^\alpha \partial \phi^\beta \,, 
\end{align} 
where $\phi^\alpha $ ($\alpha=1,...,17$) describe the scalar dof's and $K^{\alpha\beta}$ is the 
inverse of the target space metric $K_{\alpha\beta}$. 
The overall factor 2 comes from the unconventional definition of the potential $V$ in (\ref{D6Lag}). 
Here and in the following, the mass eigenvalues will be given in $g=2$ units for Mkw vacua, 
whereas for (A)dS vacua we normalize by the absolute value of the cosmological constant $\Lambda=\frac 12 V|_{\phi_0}$ .

\section{Orientifold compactifications}
\label{sec:orientifold}

In this section we study all the possible compactifications on twisted tori of type II/M-theory with O$p$-planes and/or D$p$-branes that give rise to 6-dimensional iia gauged supergravities. According to our Figure \ref{fig:branes-chart}, we need to study the following different (and inequivalent) cases:
\begin{itemize}
\item type IIB with O5/D5, O7/D7 or O9/D9,
\item (massive) type IIA with O6/D6, O8/D8, or KKO5/KK5,
\item M-theory with KKO6/KK6, 
\end{itemize}
thus making a total of 7 cases.

For each case, we will systematically analyze the configuration of the local source and the truncation of the type II/M-theory fields w.r.t. its induced involution and, where needed, the extra $\bbZ_2$ projection given by the combination of the fermionic number $(-1)^{F_L}$ and the world-sheet parity $\Omega_p$ \cite{Bergshoeff:2001pv}. Upon counting the moduli and the fluxes that survive the truncation, we will establish two  mappings: (i) the relation between the scalar fields arising from the compactification and the 6-dimensional gauged supergravity ones, and (ii) the dictionary between the background fluxes entering the compactification and the deformation parameters of the 6-dimensional supergravity sitting in the embedding tensor. 

Subsequently, by using such mappings, we will fully match the scalar potential arising from the compactification of type II or M-theory, eqs. \eqref{eq:II-potential} and \eqref{eq:11d-potential} respectively, with the scalar potential of the gauged supergravity as written in \eqref{eq:sugra-potential}. This will enable us to carry out a systematic study of vacua solutions for each of the 7 cases mentioned above.

\subsection{Massive type IIA with O6/D6}

Let us start with the class of effective theories obtained by compactifying massive type IIA supergravity on a twisted torus with one single O6-plane placed as follows:
\begin{align}
\text{O6}: 
\qquad 
\underbrace{\times|\times\times\times\times\times}_{6\mathrm{D}} \ 
\underbrace{\times - - -}_{4\mathrm{D}}
\ ,
\end{align}
which defines the following orientifold involution
\begin{align}
\sigma_{\text{O6}}:\ 
\left\{
\begin{array}{cccl}
y^0 &\mapsto & y^0 &,
\\
y^i &\mapsto & -y^i &, \ \ i=1,\, 2,\, 3\, \ .
\end{array}
\right.
\label{eq:O6-involution}
\end{align}

\subsubsection*{Fluxes and moduli}

The $\sigma_{\text{O}6}$ involution breaks $\SL(4,\bbR)$ covariance into $\mathbb{R}^+\times\SL(3,\bbR)$.
The fundamental representation of $\SL(4,\bbR)$, under which coordinates transform, branches as
\begin{align}
\mathbf{4}
\quad
\longrightarrow
\quad 
\mathbf{1}_{(+3)}\ \  \oplus\ \  \xcancel{\mathbf{3}_{(-1)}}
\ ,
\label{eq:O6-split-fundamental}
\end{align}
where all the crossed irrep's are those ones being projected out by the combination of the orientifold involution $\sigma_{\text{O}6}$, fermionic number $(-1)^{F_L}$ and world-sheet parity $\Omega_p$. The decomposition of the (physical and unphysical) scalars reads
\begin{align}
\mathbf{15} \quad \longrightarrow\quad
\mathbf{1}_{(0)}
\ \ \oplus\ \ 
\xcancel{\mathbf{3}_{(-4)}}
\ \ \oplus\ \ 
\xcancel{\mathbf{3}'_{(+4)}}
\ \ \oplus\ \ 
\mathbf{8}_{(0)}
\ .
\label{eq:O6-scalar-decomp}
\end{align}
As for the fluxes, we find
\begin{align}
\begin{array}{lclc}
H_{mnp}\in\mathbf{4'} = {\tiny\Yvcentermath1 \yng(1,1,1)}& \longrightarrow & \irrep{1}{-3}\ \ \oplus\ \ \xcancel{\irrep{3'}{+1}}&,
\\[5pt]
\omega_{mn}{}^p\in\mathbf{20} = {\tiny\Yvcentermath1 \yng(2,1)}& \longrightarrow & \irrep{3}{-1}\ \ \oplus\ \ \xcancel{\irrep{3'}{-5}}\ \ \oplus\ \ \irrep{8}{+3}\ \ \oplus\ \ \xcancel{\irrep{6'}{-1}}&,
\\[5pt]
F_{(0)}\in\mathbf{1} & \longrightarrow & \irrep{1}{0}&,
\\[5pt]
F_{mn}\in\mathbf{6} = {\tiny\Yvcentermath1 \yng(1,1)}& \longrightarrow & \xcancel{\irrep{3}{+2}}\ \ \oplus\ \ \irrep{3'}{-2}&,
\\[5pt]
F_{mnpq}\in\mathbf{1} & \longrightarrow & \xcancel{\irrep{1}{0}}&.
\end{array}
\label{eq:O6-fluxes-decomp}
\end{align}

The decomposition \eqref{eq:O6-scalar-decomp} implies that the rest of the non-universal moduli which are consistent with the orientifold involution can be parameterized by the following $\cM$ matrix
\begin{align}
\cM_{mn}
=
\left(\begin{array}{c|c}
\sigma^3 &
\\
\hline
& \sigma^{-1} M_{ij}
\end{array}
\right)
\, ,
\end{align}
where $M_{ij}$ parameterizes the $\SL(3,\bbR)/\SO(3)$ coset. Explicit parameterizations thereof can be found, \emph{e.g.}, in \cite{Bergshoeff:2003ri,deRoo:2011fa}. 
Moreover, decompositions \eqref{eq:O6-fluxes-decomp} imply that only the following flux components are non-zero 
(see Table \ref{tab:O6-dict})
\begin{align}
\begin{array}{lclclc}
H_{ijk}\equiv h\epsilon_{ijk} &,\quad &\omega_{ij}{}^0 \equiv\theta_{ij}&,\quad &\omega_{0i}{}^j\equiv\kappa_i{}^j&,
\\[5pt]
F_{(0)}\equiv f_0 &,\quad &F_{0i} \equiv f_{i}&,\quad &\text{no } F_{(4)} \text{ flux}&,
\end{array}
\end{align}
with $\theta_{ij}=-\theta_{ji}$ and $\kappa_i{}^i =0$. In what follows, 
we denote $\theta^i\equiv\frac 12 \epsilon^{ijk}\theta_{jk}$.

Summarizing, we have a set of 16 fluxes ($1+3+8+1+3$) which induce a scalar potential for $8$ scalars in total ($2+1+5$). 
The full scalar potential reads
\begin{align}
V=V_H+V_\omega+V_{F_0}+V_{F_2}+V_{\text{O}6/\text{D}6}
\ ,
\label{eq:O6-potential}
\end{align}
where tadpole cancellation requires $T_6\equiv N_{\text D6}-2N_{\text O 6}\overset{!}{=}f_0h-\theta^i f_i$.

\begin{table}[!t]
\centering
\begin{tabular}{|c|c|c|c|c|}
\hline
IIA Flux type & Flux parameters & $\sigma_{\text{O6}}$ & $(-1)^{F_L}\Omega_p$ & $\Theta$ components
\\
\hline\hline
$F_{(0)}$		& $F_{(0)}\ =\ f_0$		& $+$ 	& $+$ 	& $f_{\bar i\bar j\bar k} \ =\ f_0\ \epsilon _{\bar i\bar j \bar k}$
\\\hline
$F_{(2)} $		& $F_{0i} \ =\ f_i$		& $-$ 	& $-$ 	& $f_{\bar 1\bar j\bar k} \ =\ f_i\ \epsilon ^{ijk}$
\\\hline\hline
$H_{(3)} $		& $H_{ijk} \ =\ h\ \epsilon _{ijk}$		& $-$ 	& $-$ 	& $\zeta_{\bar 1} \ =\ h$
\\\hline
\multirow{2}{*}{$\omega$} 					& $\omega_{ij}{}^0\ =\ \theta_{ij} \equiv \epsilon_{ijk}\ \theta^k$ 	& $+$ 	& $+$ & $\zeta_{\bar i}\ =\ \theta^i$
\\
\cline{2-5}
& $\omega_{0i}{}^j \ =\ \kappa_i{}^j $ 			& $+$ 	& $+$ 	& $f_{\bar 1 i\bar j}\ =\ \kappa_i{}^j$
\\\hline
\end{tabular}
\caption{\textit{
The explicit dictionary between type IIA fluxes consistent with the O6 involution
and deformation parameters of $\cN = (1, 1)$ supergravity in six dimensions.
}}
\label{tab:O6-dict}
\end{table}

\subsubsection*{Scalar sector and fluxes/embedding tensor dictionary}

\begin{table}[!t]
\centering
\begin{tabular}{|c|c|c|c|}
\hline
IIA fields & $\sigma_{\text{O6}}$ & $(-1)^{F_L}\Omega_p$ & \# physical dof's
\\
\hline\hline
$e^0{}_0 \oplus e^i{}_j$	& $+$ 	& $+$ 	& $1+9-3=7$
\\\hline
$B_{0i} $ 					& $-$ 	& $-$ 	& $3$
\\\hline
$\Phi $ 					& $+$ 	& $+$ 	& $1$
\\\hline\hline
$C_{i} $ 					& $-$ 	& $-$ 	& $3$
\\\hline
$C_{0ij} $ 					& $+$ 	& $+$ 	& $3$
\\\hline
\end{tabular}
\caption{\textit{
The counting of the total amount of O6 allowed propagating scalar dof's in type IIA compactifications down to six dimensions. The complete set of moduli counts 17 dof's which is exactly the dimension of the supergravity coset given in \eqref{eq:sugra-coset}. Note that one needs to subtract from $e^i{}_ j$ the 3 unphysical directions corresponding to $\SO(3)$ generators in order to get the correct counting.
}}
\label{tab:O6-scalars}
\end{table}

Let us now explain how the moduli arising from the type IIA compactification described in Section \ref{sec:II-compactification} are embedded inside the scalar coset of $\cN=(1,1)$ supergravity introduced in \eqref{eq:sugra-coset}. First of all, let us count the number of propagating scalar dof's. From the internal components of the following IIA fields
\begin{align}
\{ \ e^m{}_n, \ B_{mn}, \ \Phi;\ C_m,\ C_{mnp} \ \} 
\ ,
\end{align}
we need to select those which are even under $\mathbb{Z}_{2}^{\mathrm{O}6}\equiv\sigma_{\mathrm{O}6}\,\Omega_p (-1)^{F_L}$. The result of this counting is presented in Table \ref{tab:O6-scalars}. The set of scalars coming from the reduction of the metric used to derive the scalar potential in Section \ref{sec:gauged-sugra} reads
\begin{align}
\left\{
\begin{array}{rcll}
\Lambda 
&=&
\tau^{-2}
&,
\\[5pt]
\Gamma 
&=& \rho^{1/2}\sigma^{1/2}
&,
\\[5pt]
\Sigma 
&=& \rho^{-1/4}\sigma^{3/4}
&,
\end{array}
\right.
\qquad
\cH_{MN}
=&
\left(
\begin{array}{c|c}
	\begin{array}{c|c}
	\Lambda \Gamma^{-3} & 0 
	\\\hline
	0 & \Lambda \Gamma M_{ij} 
	\end{array}
	& 0
\\\hline
0 & 
\begin{array}{c|c}
\Lambda^{-1} \Gamma^{3} & 0
\\
\hline
0 & \Lambda^{-1} \Gamma^{-1} M^{ij}
\end{array}
\end{array}
\right)
\ .
\label{eq:O6-scalars}
\end{align}

On the other hand, the embedding tensor irrep's sourced by $\zeta_M$ and $f_{M N P}$ respectively branch  w.r.t. $(\bbR^+ )^3\times \SL(3,\bbR)\subset \bbR^+\times\SO(4,4)$ as follows:
\begin{align}
\begin{array}{lcll}
\mathbf{8}_c^{(+3)} 
&\rightarrow &
\mathbf{1}
\ \oplus \
\mathbf{3}
\ \oplus \
\mathbf{1}
\ \oplus \
\mathbf{3'}
&,
\\[5pt]
\mathbf{56}_c^{(-1)} 
&\rightarrow &
\mathbf{6}
\ \oplus \
\mathbf{6'}
\ \oplus \
2\times\ (\mathbf{1}
\ \oplus \
\mathbf{3}
\ \oplus \
\mathbf{3'}
\ \oplus \
\mathbf{8} )
&.
\end{array}
\end{align}
Adopting the following splitting for $\SO(4, 4)$ light-cone coordinates
\begin{align}
M \quad \longrightarrow\quad (1,i,\bar 1,\bar i)\ ,
\label{eq:O6-split-SO44}
\end{align}
we can write down the explicit dictionary between some embedding tensor components and type IIA fluxes, thus identifying the subset of consistent deformations which admit a higher- dimensional origin. The results are collected in Table \ref{tab:O6-dict}. 

\begin{table}[!t]
\centering
\begin{tabular}{|rc|rcc|}
\hline
\multicolumn{2}{|c|}{$\cN=(1,1)$ QC/tadpoles} & \multicolumn{3}{c|}{Sources}
\\
\hline\hline
$\theta^i\ \kappa_i{}^j \stackrel{!}{=}0 $		& (Jacobi)	& KKO5/KK5: 	& $\stackrel{0}{-} \ \stackrel{i}{-} \ \stackrel{j}{\text{ISO}} \ \stackrel{k}{-}$ 	& $(\times 3)$
\\\hline\hline
\multicolumn{2}{|c|}{$\cN=(2,2)$ QC/tadpoles} & \multicolumn{3}{c|}{Sources}
\\
\hline\hline
$f_0\ h- \theta^i \ f_i \stackrel{!}{=}0 $		& (BI $C_{(1)}$)	& O6/D6: 	& $\stackrel{0}{\times} \ \stackrel{i}{-} \ \stackrel{j}{-} \ \stackrel{k}{-}$ 	& $(\times 1)$
\\\hline
\end{tabular}
\caption{\textit{
Non-vanishing QC \eqref{eq:QC}, \eqref{eq:QC-extra} and their higher-dimensional origin for the flux compactification of massive type IIA with O6/D6 given in Table \ref{tab:O6-dict}, where BI stands for Bianchi identities and Jacobi refers to the condition in \eqref{eq:II-Jacobi}. A description of the QC as restrictions for the existence of additional local sources is given.
}}
\label{tab:O6-tadpoles}
\end{table}

Using the dictionary presented in Table \ref{tab:O6-dict}, if we restrict the embedding tensor to those components corresponding to IIA fluxes, the QC in \eqref{eq:QC} reduce to 
\begin{align}
\theta^i \kappa_i{}^j=0 
\ ,
\end{align}
which correspond to the Jacobi identities of the underlying group manifold already found in \eqref{eq:II-Jacobi}. These can be interpreted as conditions for the absence of KK monopoles \cite{Villadoro:2007yq}, which would further break supersymmetry down to eight supercharges. Furthermore, as a cross-check, one can derive the form of the extra QC \eqref{eq:QC-extra} required to have a maximal supergravity description for a gauging arising from a type IIA compactification. We find that they correspond to the absence of O6/D6 sources, \emph{i.e.},
\begin{align}
T_6\ \equiv\ f_0 h- \theta^i f_i = 0
\ .
\end{align}
Further details on the physical interpretation of these constraints are given in Table \ref{tab:O6-tadpoles}.

By inserting the parameterization of the scalars given in \eqref{eq:O6-scalars} together with the embedding tensor/fluxes dictionary of Table \ref{tab:O6-dict} inside the supergravity potential \eqref{eq:sugra-potential}, we exactly reproduce the moduli potential computed in \eqref{eq:O6-potential} from dimensional reduction upon fixing the gauge coupling to $g = 2$.

\subsubsection*{Critical points}

\begin{table}[!t]
\centering
\begin{tabular}{|c|c|c|c|c|c|c|c|}
\hline
Sol \# & $f_0$ & $f_i$ & $	h$ & $\kappa_i{}^j$ & $\theta^i$ & $T_6$ & $m^2$
\\\hline\hline
1 	&	$\alpha$ 	& $\beta_i$	& $\alpha$ 		&$\mathbb{O}_3$	& $\beta_i$ &
${\alpha^2-|\vec \beta|^2}$ &
\begin{tabular}{c}
$0_{(\times 13)}$, {$(\alpha^2 +|\vec \beta|^2)_{(\times 3)}$},\\
 {$4(\alpha^2 +|\vec \beta|^2)_{(\times 1)}$}
 \end{tabular}
\\\hline
2 	&	$\alpha$ 	& $0$	& $\alpha$ 		& 
$\begin{pmatrix}
0 & \beta_1 & \beta_2
\\
-\beta_1 & 0 & \beta_3
\\
-\beta_2 & -\beta_3 & 0
\end{pmatrix}$			& $0$  
& $\alpha^2$&
\begin{tabular}{c}
$0_{(\times 9)}$, $\alpha^2_{(\times 1)}$, {$4\alpha^2_{(\times 1)}$}, 
{$|\vec\beta|^2_{(\times 2)}$},  \\
{$4|\vec\beta|^2_{(\times 2)}$, $(\alpha^2+|\vec\beta|^2)_{(\times 2)}$ }
\end{tabular}
\\\hline
3 	&	$0$ 	& $0$	& $0$ 		& 
$\begin{pmatrix}
0 & \alpha & \beta
\\
0 & 0 & 0
\\
0 & 0 & 0
\end{pmatrix}$			& $\begin{pmatrix} 0\\ -\beta \\ \alpha\end{pmatrix}$  & 0&
\begin{tabular}{c}
$0_{(\times 13)}$, $(\alpha-\beta)^2_{(\times 3)}$, \\
$4(\alpha-\beta)^2_{(\times 1)}$\end{tabular}
\\\hline
\end{tabular}
\caption{\textit{
Critical points of the scalar potential induced by the compactification of mIIA with O6/D6. In this case all of the above solutions are Mkw. The solutions can be embedded into $\cN=(2,2)$ theory iff $T_6$ vanishes. 
Mass eigenvalues are computed in $g=2 $ units. We use the notation $\vec\beta\equiv(\beta_1,\beta_2,\beta_3)$.
}}
\label{tab:O6-solutions}
\end{table}

Establishing an embedding tensor/fluxes dictionary enables us to study the critical points of the theory in a systematic way. By applying the going-to-the-origin (GTTO) method \cite{Dibitetto:2011gm}, we scan the embedding tensor configurations that allow for critical points in the potential when the scalar fields take the values at the origin of the scalar manifold. This amounts to solving a set of quadratic equations, in terms of the embedding tensor components.

It is worthwhile to stress that, despite that only a subset of the scalar fields of half-maximal gauged supergravity appear as deformations of the 10D metric (and hence in the moduli potentials in \eqref{eq:II-potential}), we must ensure that the equations of motion of all the scalar dof's are satisfied, including those modes that appear in the reduction Ansatz of the $p$-form potentials which we omitted for simplicity.
This is strictly necessary in order to have a consistent vacuum solution.  In this respect, the formulation of the effective theory as a gauged supergravity simplifies the problem.

The consistency of the compactification of mIIA with O6/D6 allows for the fluxes given in Table \ref{tab:O6-dict}. When considering the scalar potential that they give rise to, we find 3 families of critical points.  In Table \ref{tab:O6-solutions} we show the embedding tensor (or, using Table \ref{tab:O6-dict}, the fluxes) configuration and  the corresponding mass eigenvalues for each case. We note the existence of a critical point for a configuration that only carries metric flux and no gauge fluxes. 
In Appendix \ref{app:global} we discuss the global properties of internal manifolds corresponding to each of the critical points.

\subsection{Massive type IIA with O8/D8}

Let us now consider the effective theory obtained when compactifying (massive) type IIA with O8/D8 planes. When an O8-plane is placed in this form:
\begin{align}
\text{O8}: 
\qquad 
\underbrace{\times|\times\times\times\times\times}_{6\mathrm{D}} \ 
\underbrace{\times \times \times -}_{4\mathrm{D}}
\ ,
\end{align}
it defines the following orientifold involution
\begin{align}
\sigma_{\text{O8}}:\ 
\left\{
\begin{array}{cccl}
y^i &\mapsto & y^i &, \ \ i=1,\, 2,\, 3\, ,
\\
y^0 &\mapsto & -y^0 &.
\end{array}
\right.
\end{align}

\subsubsection*{Fluxes and moduli}

The involution $\sigma_{\text{O}8}$ breaks $\SL(4,\bbR)$ covariance into $\bbR^+\times \SL(3,\bbR)$  and the fundamental and adjoint representations split as in \eqref{eq:O6-split-fundamental} and \eqref{eq:O6-scalar-decomp}, respectively. According to the $\bbZ_2$ truncation induced by $\sigma_{\text{O}8}$ and the world-sheet parity $\Omega_p$ \cite{Bergshoeff:2001pv}, the surviving fluxes are given in Table \ref{tab:O8-dict}, and they correspond to
\begin{align}
\begin{array}{rcllrcllrcll}
\omega_{ij}{}^k&\equiv&{\frac 12}\theta_{[i}\delta_{j]}^k+\epsilon _{ijl} \kappa^{(lk)}&,&	\omega_{i0}{}^0&\equiv&-\theta_i	&,&		H_{0ij}&\equiv&\epsilon _{ijk}\ h^k&,
\\[5pt]
F_{ij}&\equiv&\epsilon _{ijk}\ f^k			&,&	F_{0ijk}&\equiv&\epsilon _{ijk} f_4			&,&		\text{no}& F_{(0)} &   \text{ flux}&.
\end{array}
\end{align}

The decomposition \eqref{eq:O6-scalar-decomp} implies that all the non-universal moduli that are consistent with the orientifold involution are embedded in the matrix $\cM$ as follows:
\begin{align}
\cM_{mn}
=
\left(\begin{array}{c|c}
\sigma^3 &
\\
\hline
& \sigma^{-1} M_{ij}
\end{array}
\right)
\, ,
\end{align}
where $M_{ij}$ parameterizes the $\SL(3,\bbR)/\SO(3)$ coset.
In summary, we have a set of 18 fluxes ($8+3+3+1+3$) and 8 scalars ($2+1+(8-3)$). Because we find $T_8=0$, the term $V_{\text{O}p/\text{D}p}$ that contributes to the scalar potential vanishes, \emph{i.e.}, there are no $\cN=(2,2)$ tadpoles\footnote{The suitable configuration of spacetime filling sources enforcing the vanishing of the corresponding flux tadpole is given by an O8 plane with 8 D8 branes on top.}.

\begin{table}[!t]
\centering
\begin{tabular}{|c|c|c|c|c|}
\hline
IIA Flux type & Flux parameters & $\sigma_{\text{O8}}$ & $\Omega_p$ & $\Theta$ components
\\
\hline\hline
$F_{(2)} $		& $F_{ij}\ = \ \epsilon _{ijk}\ f^k$		& $+$ 	& $+$ 	& $f_{0ij}\ = \ \epsilon _{ijk} \ f^k$
\\\hline
$F_{(4)} $		& $F_{0ijk}\ =\ \epsilon _{ijk}\ f_4$		& $-$ 	& $-$ 	& $f_{ijk}\ =\ \epsilon _{ijk}\ f_4$
\\\hline\hline
$H_{(3)} $		& $H_{0ij}\ =\ \epsilon _{ijk}\ h^k$		& $-$ 	& $-$ 	& ${f_{\bar0 ij}\ = \ \epsilon _{ijk}\ h^k}$
\\\hline
\multirow{2}{*}{$\omega$} 					& $\omega_{ij}{}^k \ = \ {\frac 12}\theta_{[i}\ \delta_{j]}^k+\epsilon _{ijl}\ \kappa^{(lk)} $ 			& $+$ 	& $+$ 	& $f_{ij\bar k}\ =\ {\frac 12}\theta_{[i}\delta_{j]}^k+\epsilon _{ijl}\ \kappa^{(lk)}$
\\
\cline{2-5}
& $\omega_{i0}{}^0 \ = \ -\theta_i $ 			& $+$ 	& $+$ 	& ${f_{i0\bar 0}=\frac 12 \theta_i}$
\\ \hline
\end{tabular}
\caption{\textit{
The explicit dictionary between type IIA fluxes consistent with the O8 involution
and deformation parameters of $\cN = (1, 1)$ supergravity in six dimensions.
}}
\label{tab:O8-dict}
\end{table}

\subsubsection*{Scalar sector and fluxes/embedding tensor dictionary}

Let us now discuss the explicit embeddings of both the moduli arising from Type IIA compactification inside the scalar coset \eqref{eq:sugra-coset} and  the fluxes inside the various embedding tensor irrep's.

\begin{table}[!t]
\centering
\begin{tabular}{|c|c|c|c|}
\hline
IIA fields & $\sigma_{\text{O8}}$ & $\Omega_p$ & \# physical dof's
\\
\hline\hline
$e^i{}_j \oplus e^0{}_0$	& $+$ 	& $+$ 	& $9-3+1=7$
\\\hline
$B_{0i} $ 					& $-$ 	& $-$ 	& $3$
\\\hline
$\Phi $ 					& $+$ 	& $+$ 	& $1$
\\\hline\hline
$C_{i} $ 					& $+$ 	& $+$ 	& $3$
\\\hline
$C_{0ij} $ 					& $-$ 	& $-$ 	& $3$
\\\hline
\end{tabular}
\caption{\textit{
Counting of the total 17 propagating scalar dof's allowed by O8-planes in type IIA compactifications down to six dimensions. This is exactly the dimension of the supergravity coset given in \eqref{eq:sugra-coset}, once we subtract from $e^i{}_ j$ the 3 unphysical directions corresponding to $\SO(3)$ generators.
}}
\label{tab:O8-scalars}
\end{table}

The set of scalar propagating dof's that survive the $\Omega_p$ projection amounts to 17 and their higher-dimensional origin is presented in Table \ref{tab:O8-scalars}. The explicit mapping between these fields and the scalar fields of iia supergravity theory is given by
\begin{align}
\left\{
\begin{array}{rcll}
\Lambda 
&=&
\tau^{-2}
& ,
\\[5pt]
\Gamma &=& \rho^{1/2}\sigma^{1/2}
& ,
\\[5pt]
\Sigma 
&=& \rho^{-1/4}\sigma^{3/4}
&,
\end{array}
\right.
\qquad
\cH_{MN}
=&
\left(
\begin{array}{c|c}
	\begin{array}{c|c}
	\Lambda \Gamma^{-3} & 0 
	\\\hline
	0 & \Lambda \Gamma M_{ij} 
	\end{array}
	& 0
\\\hline
0 & 
\begin{array}{c|c}
\Lambda^{-1} \Gamma^{3} & 0
\\
\hline
0 & \Lambda^{-1} \Gamma^{-1} M^{ij}
\end{array}
\end{array}
\right)
\ .
\end{align}

On the other hand, the mapping between the fluxes that survive the O8 truncation and the $\SL(3,\bbR)$-irrep's of the embedding tensor is given in Table \ref{tab:O8-dict}, where we have used the notation for the splitting of the $\SO(4,4)$ light-cone coordinates introduced in \eqref{eq:O6-split-SO44}. If we apply this dictionary to the embedding tensor  and the previous mapping to the supergravity scalars in \eqref{eq:sugra-potential}, we automatically obtain the potential \eqref{eq:II-potential}.

\begin{table}[!t]
\centering
\begin{tabular}{|rc|rcc|}
\hline
\multicolumn{2}{|c|}{$\cN=(1,1)$ QC/tadpoles} & \multicolumn{3}{c|}{Sources}
\\
\hline\hline
$f^i\theta_i\stackrel{!}{=}0 $		& (BI $C_{(1)}$)	& D6/O6: 	& $\stackrel{0}{\times} \ \stackrel{i}{-} \ \stackrel{j}{-} \ \stackrel{k}{-}$ 	& $(\times 1)$
\\\hline
$\theta_i\kappa^{ij}\stackrel{!}{=}0 $		& (Jacobi)	& KK5/KKO5: 	& $\stackrel{0}{-} \ \stackrel{i}{-} \ \stackrel{j}{\text{ISO}} \ \stackrel{k}{-}$ 	& $(\times 3)$
\\\hline
\end{tabular}
\caption{\textit{
Non-vanishing QC \eqref{eq:QC}, \eqref{eq:QC-extra} and their higher-dimensional origin for the flux compactification of type mIIA with O8/D8 given in Table \ref{tab:O8-dict}, where BI stands for Bianchi identities and Jacobi refers to the condition \eqref{eq:II-Jacobi}. A description of the QC as restrictions for the existence of additional local sources is given.
}}
\label{tab:O8-tadpoles}
\end{table}

Using the embedding tensor/fluxes dictionary, we can study the QC \eqref{eq:QC} that survive when we restrict ourselves to the fluxes of Table \ref{tab:O8-dict}. In Table \ref{tab:O8-tadpoles} we show the set of non-vanishing constraints and their physical interpretation. In particular, such conditions can be understood as the $\cN=(1,1)$ tadpoles and impose the absence of the various undesired supersymmetry breaking sources which appear in the table.

\subsubsection*{Critical points}

\begin{table}[!t]
\centering
\begin{tabular}{|c|c|c|c|c|c|c|}
\hline
Sol \# & $f_4$ & $f^i$ & $	h^i$ & $\theta_i$ & $\kappa^{ij}$ & $m^2$
\\\hline\hline
1 	&	$0$ 	& $0$	& $0$ 		&$0$			& $\text{diag}(\alpha,\alpha,0)$
& $0_{(\times 11)}$, $\alpha^2_{(\times 4)}$, $4\alpha^2_{(\times 2)}$
\\\hline
\end{tabular}
\caption{\textit{
Critical points of the scalar potential induced by the compactification of mIIA with O8/D8. In this case, we obtain a 1-parameter family of solutions, which induces a Minkowski-type universe.
}}
\label{tab:O8-solutions}
\end{table}
We are now ready to study the critical points of the scalar potential when the non-vanishing embedding tensor components are the ones given in Table \ref{tab:O8-dict}. 
When we solve the equations of motion of the scalar fields and the QC \eqref{eq:QC}, we obtain a unique 1-parameter family of solutions, which corresponds to a Mkw vacuum. Further details are given in Table \ref{tab:O8-solutions}. 
In Appendix \ref{app:global} we show that this solution can be obtained as a compactification on a globally well defined twisted torus.

\subsection{Massive type IIA with KKO5/KK5}

We will focus on the class of effective theories obtained by compactifying type IIA supergravity with one single KKO5-plane placed as follows:	
\begin{align}
\text{KKO5}: 
\qquad 
\underbrace{\times|\times\times\times\times\times}_{6\mathrm{D}} \ 
\underbrace{\text{ISO} - - - }_{4\mathrm{D}}
\ ,
\end{align}
which defines the following orientifold involution
\begin{align}
\sigma_{\text{KKO5}}:\ \ 
y^m \quad \mapsto \quad -y^m &, \ \ m=0,\, i,\, j,\, k\, .
\end{align}
We have split the $\SL(4,\bbR)$ index as $m=(0,i)$, $i=1,2,3$ and denoted the isometry direction as $y^0$. It is perhaps worth mentioning that this case stands out w.r.t. all the others treated in this work. Turning on metric flux $\omega_{mn}{}^p$ is not allowed due to parity arguments. However, in this particular setup, a sphere reduction turns out to be consistent. An explicit evidence for this is provided by the supersymmetric $\mathrm{AdS}_6\times S^4/\bbZ_k$ vacuum originally constructed in \cite{Brandhuber:1999np} as near horizon limit of a D4 -- D8 -- KK5 brane system. 
The underlying gauged supergravity has gauge group $\mathrm{ISO}(3)$ gauge group, and the embedding tensor is associated with the extrinsic curvature of $S^4$ \cite{Dibitetto:2018iar}.

\subsubsection*{Fluxes and moduli}

In this case, the involution generated by the KKO5 orientifold as a local source preserves the $\SL(4,\bbR)$ covariance. In addition, this BPS object does not impose any additional $\bbZ_2$ truncation \cite{Bergshoeff:2001pv}.
\begin{table}[!t]
\centering
\begin{tabular}{|c|c|c|c|c|}
\hline
IIA Flux type & Flux parameters & $\sigma_{\text{KKO5}}$ & no extra $\bbZ_2$ & $\Theta$ components
\\
\hline\hline
$F_{(0)}$		& $F_{(0)}\ =\ f_0$		& $+$ 	& $+$ 	& $\zeta_{\bar0}\ =\ f_0$
\\\hline
$F_{(4)} $		& $F_{0ijk}=f_4\ \epsilon _{ijk}$		& $+$ 	& $+$ 	& $\zeta_0=f_4$
\\\hline
$F_{(2)} $		& $F_{mn}=f_{mn}$		& $+$ 	& $+$ 	& $\zeta_A=\frac12 f_{mn}[G_A]^{mn}$
\\\hline\hline
$K $	& $K_{(mn)}$		& $+$ 	& $+$ 	& $\tilde Q_{ij}\ \oplus\ Q^{00}$
\\\hline
\end{tabular}
\caption{\textit{
The explicit dictionary between type IIA fluxes consistent with the KKO5 involution
and deformation parameters of $\cN = (1, 1)$ supergravity in six dimensions. The $K_{mn}$ tensor denotes the extrinsic curvature of the 4-sphere as explained in \cite{Danielsson:2015tsa}.
}}
\label{tab:KKO5-dict}
\end{table}
In Table \ref{tab:KKO5-dict} we show in detail the set of fluxes that are compatible with the KKO5 orientifold projection. Explicitly, it consists of
\begin{align}
\begin{array}{rcllrcll}
K_{mn}&\equiv & K_{(mn)} &,\qquad& F_{(0)}& \equiv&f_0 &,
\\[5pt]
F_{mn}&\equiv &f_{mn} &,\qquad& F_{0ijk}&\equiv& f_4\ \epsilon _{ijk} &,
\end{array}
\end{align}
where $K_{(mn)}$ denotes the extrinsic curvature of the $S^4$ \cite{Danielsson:2015tsa}. 
 
In summary, we have 11 scalars ($1+1+(15-6)$) and 18 fluxes ($10+1+6+1$). Regarding the term $V_{\text{KKO}5/\text{KK}5}$ in the potential, in the next section we will explain its non-trivial contribution, due to the existence of an $\cN=(2,2)$ tadpole.

\subsubsection*{Scalar sector and fluxes/embedding tensor dictionary}

\begin{table}[!t]
\centering
\begin{tabular}{|c|c|c|c|}
\hline
IIA fields & $\sigma_{\text{KKO5}}$ & no extra $\bbZ_2$ & \# physical dof's
\\
\hline\hline
$e^m{}_n$	& $+$ 	& $+$ 	& $16-6=10$
\\\hline
$B_{mn} $ 					& $+$ 	& $+$ 	& $6$
\\\hline
$\Phi $ 					& $+$ 	& $+$ 	& $1$
\\\hline\hline
$C_{(1)} $ 					& $-$ 	& $+$ 	& $-$
\\\hline
$C_{(3)} $ 					& $-$ 	& $+$ 	& $-$
\\\hline
\end{tabular}
\caption{\textit{Counting of the total 17 propagating scalar dof's allowed by KKO5-planes in type IIA compactifications down to six dimensions. This is exactly the dimension of the supergravity coset given in \eqref{eq:sugra-coset}, once we subtract from $e^m{}_n$ the 6 unphysical directions corresponding to $\SO(4)$ generators.
}}
\label{tab:KKO5-scalars}
\end{table}
In Table \ref{tab:KKO5-scalars} we show the origin of the full set of scalar fields from the type IIA field contents. However, to obtain the relation between these fields and the supergravity scalars, we need to do a previous consideration. In this particular situation, the embedding of the four compact internal directions turns out to be \emph{spinorial}, which is possible due to the presence of a triality of $\SO(4,4)$ irrep's of dimension $8$. The isomorphism $\mathfrak{sl}(4,\bbR)\cong \mathfrak{so}(3,3)$ allows us to construct a specific mapping between the adjoint of $\SL(4,\bbR)$ and the fundamental of $\SO(3,3)$. To do so, it is convenient to split the $\SO(4,4)$ light-cone coordinates as
\begin{align}
M \qquad 
\rightarrow\qquad
(0,i,\bar 0,\bar i)
\ ,
\qquad\qquad
A\equiv(i,\bar i)
\ ,
\label{eq:SO44-to-SO33}
\end{align}
where $i=1,2,3$ is an $\SO(3)$ index and $A=1,2,3,\bar 1,\bar 2,\bar 3$ is an $\SO(3,3)$ index expressed in the light-cone basis. Then, the mapping of a vector $V_A$ of $\SO(3,3)$ to a 2-form $v_{mn}$ of $\SL(4,\bbR)$ is
\begin{align}
V_A=\frac12\,[G_A]^{mn} v_{mn} \ ,
\end{align}
where the set of matrices $[G_A]^{mn}$ are the so-called 't Hooft symbols, which explicitly realize the aforementioned map. Further properties and conventions concerning this map can be found in Appendices of \cite{Dibitetto:2012rk,Dibitetto:2015bia}. 

Then, the mapping relating the propagating scalars that arise from type IIA compactification with KKO5 planes and the gauged supergravity fields is given by
\begin{align}
\left\{
\begin{array}{rcll}
\Lambda 
&=&
\rho^{2}
& ,
\\[5pt]
\Sigma 
&=& 
\tau^{-1}
&,
\end{array}
\right.
\qquad
\cH_{MN}
=&
\left(
\begin{array}{c|c}
	\begin{array}{c|c}
	\Lambda  & 0 
	\\\hline
	0 & {\cM_{ij} }
	\end{array}
	& 0
\\\hline
0 & 
\begin{array}{c|c}
\Lambda^{-1}  & 0
\\
\hline
0 & {\cM^{\bar i\bar j}}
\end{array}
\end{array}
\right)
\ ,
\end{align}
where $\cM_{AB}$ is given by
\begin{align}
\cM_{AB}
=
\frac12 [G_A]^{mp}[G_B]^{nq} \cM_{mn} \cM_{pq}
\ .
\label{eq:SO33-scalars}
\end{align}
In this case, the expression of the vielbein $\cV_A{}^{\underline{I\hat J}}$ which squares to $\cM_{AB}$ is given by
\begin{align}
\cV_A{}^{\underline{I}\underline{\hat J}}
=
\frac{1}{4\sqrt2} \cV_m{}^{\underline{m}} \cV_n{}^{\underline{n}} [G_A]^{mn} [\Gamma_{\underline{m}}]^{\alpha\hat\beta} [\bar \Gamma_{\underline{n}}]^{\hat\delta\gamma}(\sigma^{\underline{I}})_\alpha{}^\gamma(\sigma^{\underline{\hat J}})_{\hat\beta}{}^{\hat\delta}
\ ,
\label{eq:SO33-vielbein}
\end{align}
where $\underline{I}$ and $\underline{\hat{I}}$ are indices of the fundamental representation of each of two factors of $\SO(3)\times\SO(3)$ and $\alpha$ and $\hat\alpha$ are indices of the adjoint representation of each of the two factors of $\SU(2)\times\SU(2)$. Finally, the mapping between the fundamental of $\SO(3)$ and the adjoint of $\SU(2)$ is given by the Pauli matrices $(\sigma^{\underline{I}})_\alpha{}^\beta$, whereas $(\Gamma_{\underline{m}})^{\alpha\hat\beta}$ are Dirac matrices in the Weyl representation (\emph{c.f.} \cite{Dibitetto:2015bia}).

The embedding tensor/fluxes dictionary will identify the consistent deformations of supergravity that arise from the compactification with KKO5-planes. The mapping is given in Table \ref{tab:KKO5-dict}. The $\SO(3,3)$  3-form of the embedding tensor $f_{ABC}\subset f_{MNP}$, which carries the metric flux written in terms of the extrinsic curvature, is parameterized as follows:
\begin{align}
f_{ABC} 
= {2}
\left(
	\frac12\ \delta_{[m}^{[r}\ Q_{n][p} \ \delta_{q]}^{s]}
	+\frac14\ \epsilon _{tmn[p} \ \tilde Q^{t[r}\ \delta_{q]}^{s]}
	\right) [G_A]^{mn}\ [G_B]^{pq}\ [G_C]_{rs}
\ ,
\end{align}
where the symmetric matrices $Q$ and $\tilde Q$ are the embedding tensor components specified in Table \ref{tab:KKO5-dict} and $G_A$ are the 't Hooft symbols.

\begin{table}[!t]
\centering
\begin{tabular}{|c|rcc|}
\hline
$\cN=(1,1)$ QC/tadpoles & \multicolumn{3}{c|}{Sources}
\\
\hline\hline
$\tilde \D F_2\stackrel{!}{=}0 $ ~~ (BI $C_{(1)}$)	& D6/O6: 	& $\stackrel{0}{-} \ \stackrel{i}{-} \ \stackrel{j}{-} \ \stackrel{k}{\times}$ 	& $(\times 3)$
\\\hline
$\tilde \D F_2\stackrel{!}{=}0$	~~~(BI $C_{(1)}$)	& D6/O6: 	& $\stackrel{0}{\times} \ \stackrel{i}{-} \ \stackrel{j}{-} \ \stackrel{k}{-}$ 	& $(\times 1)$
\\\hline\hline
$\cN=(2,2)$ QC/tadpoles & \multicolumn{3}{c|}{Sources}
\\
\hline\hline
$f_0Q_{00}\stackrel{!}{=}0\,,\  f_4 \tilde Q_{ij}\stackrel{!}{=}0 $	& KK5/KKO5: 	& $\stackrel{0}{\text{ISO}} \ \stackrel{i}{-} \ \stackrel{j}{-} \ \stackrel{k}{-}$ 	& $(\times 1)$
\\\hline
\end{tabular}
\caption{\textit{
Non-vanishing QC \eqref{eq:QC}, \eqref{eq:QC-extra} and their higher-dimensional origin for the flux compactification of type mIIA with KKO5/KK5 given in Table \ref{tab:KKO5-dict}, where BI stands for Bianchi identities and 
$\tilde \D \equiv \D+\omega\ \wedge \ $. A description of the QC as restrictions for the existence of additional local sources is given.
}}
\label{tab:KKO5-tadpoles}
\end{table}

If we study the QC \eqref{eq:QC} by restricting ourselves to the above configuration of fluxes, we obtain some surviving conditions. These conditions can be interpreted as restrictions for the presence of additional sources and correspond to the $\cN=(1,1)$ tadpoles written in Table \ref{tab:KKO5-tadpoles}.
Regarding the extra QC \eqref{eq:QC-extra}, they correspond to an $\cN=(2,2)$ tadpole, which is shown in Table \ref{tab:KKO5-tadpoles}. The value of $T_{5,1}$ is given by
\begin{align}
T_{5,1}=f_{0}\ Q_{00}
\ .
\end{align}

Then, upon using the parameterization of the fluxes and the scalar fields and choosing $T_{{5,1}}=f_0\ Q_{00}$, both the gauged supergravity potential \eqref{eq:sugra-potential} and the potential from the dimensional reduction \eqref{eq:II-potential} are unambiguously identified.

\subsubsection*{Critical points}

\begin{table}[!t]
\centering
\resizebox{\textwidth}{!}{%
\begin{tabular}{|c|c|c|c|c|c|}
\hline
Sol \# & $\tilde Q^{mn}$ & $f_0$ & $f_4$ & $f_{mn}$ & $Q_{mn}$ 
\\\hline\hline
1 	&	$\text{diag}(0,3\alpha,3\alpha,3\alpha)$ 	& $-\alpha$	& $\alpha$ 		&$0$			& $\text{diag}(3\alpha,0,0,0)$ 
\\\hline
2 	&	$\text{diag}(0,\alpha,\alpha,\alpha)$ 	& $-\alpha$	& $\alpha$ 		&$0$			& $\text{diag}(\alpha,0,0,0)$ 
\\\hline
3 (3$'$) 	&	$\mathbb O_4$ 	& $\alpha$ ($0$)	& $0$ ($\alpha$) 		&$0$			& $\text{diag}(\pm\alpha,0,0,0)$ 
\\\hline
4 	&	$(-5+2\sqrt7)\ \text{diag}(0,\alpha,\alpha,\alpha)$ 	& $\alpha$	& $-\alpha$ 		&$0$			& $(-7+2\sqrt7)\ \text{diag}(\alpha,0,0,0)$ 
\\\hline
5 	&	$(5+2\sqrt7)\ \text{diag}(0,\alpha,\alpha,\alpha)$ 	& $-\alpha$	& $\alpha$ 		&$0$			& $-(7+2\sqrt7)\ \text{diag}(\alpha,0,0,0)$ 
\\\hline
{6} 	&	$\frac{\alpha}{\alpha+\beta} \text{diag}(-\alpha,\beta,\beta,\beta)$ 	& $\alpha$	& $\beta$ 		&$0$	&
 $\frac{\beta}{\alpha+\beta} \text{diag}(-\beta,\alpha,\alpha,\alpha)$ 
\\\hline
{7} 	&	$\alpha\ \text{diag}(-1, 7\pm4\sqrt 3,7\pm4\sqrt 3,7\pm4\sqrt 3)$ 	& $0$	& $0$ 		&$0$	&
 $\alpha \ \text{diag}(-(7\pm 4\sqrt 3),1,1,1)$ 
\\\hline
\end{tabular}}
\caption{\textit{
Critical points of the scalar potential induced by the compactification of mIIA with KKO5/KK5. Solution 1 is the supersymmetric AdS vacuum found in \protect\cite{Brandhuber:1999np}. 
}
}
\label{tab:KKO5-solutions}
\end{table}

\begin{table}[!t]
\centering
\begin{tabular}{|c|c|c|}
\hline
Sol \# &  $\Lambda=\frac 12 V_0$ & $m^2$
\\\hline\hline
1 	&  {$-\frac 52 g^2\alpha^2$} & $\frac{12}{5} _{(\times 1)}$,$\frac{7}{5} _{(\times 3)}$, 
$-\frac{3}{5} _{(\times 4)}$, $\frac{3}{5} _{(\times 5)}$, $-\frac{2}{5} _{(\times 1)}$, 
$0_{(\times 3)}$
\\\hline
2 	&	{$-\frac 12 g^2\alpha^2 $} & $2 _{(\times 1)}$, $-1 _{(\times 8)}$, 
$1 _{(\times 4)}$, $0 _{(\times 4)}$
\\\hline
3 (3$'$) 	& $0$ & $4\alpha^2_{(\times 1)}$, $\alpha^2_{(\times 3)}$, $0 _{(\times 13)}$
\\\hline
4 	& {$\frac 12\left(8 \sqrt{7}-21\right) g^2\alpha^2$} & 
\begin{tabular}{c}$\frac 1{14}(77+36\sqrt 7\pm \sqrt{22057+7896\sqrt 7})_{(\times 1)}$, 
$5+\frac{12}{\sqrt 7}_{(\times 3)}$, \\ $1+\frac{8}{\sqrt 7}_{(\times 5)}$, 
$2+\frac{4}{\sqrt 7}_{(\times 1)}$, $-1+\frac{4}{\sqrt 7}_{(\times 3)}$, $0_{(\times 3)}$
\end{tabular}
\\\hline
5 	 & {$-\frac 12\left(8 \sqrt{7}+21\right)g^2 \alpha^2$} & 
\begin{tabular}{c}$\frac 1{14}(-77+36\sqrt 7\pm \sqrt{22057-7896\sqrt 7})_{(\times 1)}$, 
$-5+\frac{12}{\sqrt 7}_{(\times 3)}$, \\ $-1+\frac{8}{\sqrt 7}_{(\times 5)}$, 
$-2+\frac{4}{\sqrt 7}_{(\times 1)}$, $1+\frac{4}{\sqrt 7}_{(\times 3)}$, $0_{(\times 3)}$
\end{tabular}
\\\hline 
{6} & $\frac 12 g^2 \alpha \beta$ & $0_{(\times 3)}$, $1_{(\times 5)}$, $\frac{(\alpha+\beta)^2\pm \sqrt{(\alpha+\beta)^4+16\alpha^2\beta^2}}{4|\alpha \beta|} _{(\times 3)}$, $\frac{\lambda_{(i)}}{2|\alpha\beta|(\alpha\beta)^3}$ ($i=1,2,3$)  \\\hline
{7} & 0 & $32(7\pm 4\sqrt 3)\alpha^2 _{(\times 9)}$, $0_{(\times 8)}$ \\\hline
\end{tabular}
\caption{\textit{
Potential and mass eigenvalues for the critical points of the scalar potential induced by the compactification of mIIA with KKO5/KK5. In this case, we obtain two families of Mkw vacua, 3 families of AdS vacua, a 1-parameter family of dS solutions {and a 2-parameter family of (A)dS solutions}. Solution 1 is the supersymmetric AdS vacuum found in \protect\cite{Brandhuber:1999np}.
When the potential is nonvanishing, the mass eigenvalues are normalized by $|\Lambda|$. 
In $D=6$ AdS spacetime, the Breitenlohner-Freedman bound reads $m^2_{\rm BF}=-\frac 58 |\Lambda|$.  
}}
\label{tab:KKO5-solutions-masses}
\end{table}
Let us consider the flux configuration given in Table \ref{tab:KKO5-dict} and  evaluate the corresponding non-vanishing embedding tensor components in the scalar potential \eqref{eq:sugra-potential}. When we solve the equations of motion of the scalar fields as well as the QC \eqref{eq:QC}, we obtain 7 families of solutions. The flux configuration for the full set of solutions is given in Table \ref{tab:KKO5-solutions}, whereas the type of vacua that they give rise to and the mass spectrum are shown in Table \ref{tab:KKO5-solutions-masses}.

Let us note that we obtain a particular solution ({Solution 1}) which precisely corresponds to the supersymmetric AdS vacuum found in\cite{Brandhuber:1999np}, where the residual SU($2$) isometries are interpreted as the R-symmetry of the dual $\cN=1$ SCFT$_5$. Other AdS vacua do not preserve any supersymmetry. A quick way to see this is to check, as a necessary condition, if the mass spectrum fulfills the Breitenlohner-Freedman bound \cite{Breitenlohner:1982bm}. 
The Breitenlohner-Freedman bound is the lowest mass eigenvalue, for which the scalar field in AdS is stable \cite{Ishibashi:2004wx}. Solution 1 satisfies this bound, whereas solutions 2 and 4 do not. 

{Solution 6 describes the 2-parameter family of AdS vacua for $\alpha\beta<0$ and the dS vacua  for $\alpha\beta>0$. }
For the mass eigenvalues of Solution 6, the roots $\lambda_{(i)}$ satisfies the following cubic equation
\begin{align}
\label{}
f(\lambda) \equiv \lambda^3-2(2\alpha^2+\alpha\beta+2\beta^2)\lambda^2+
2\alpha\beta(\alpha^2+6\alpha\beta+\beta^2)\lambda +12 \alpha^2\beta^2(\alpha+\beta)^2=0 \,. 
\end{align}
On account of $f(0)=12(\alpha\beta)^2(\alpha+\beta)^2>0$ and $f(-2\alpha\beta)=-8(\alpha\beta)^2(\alpha+\beta)^2<0$, 
the corresponding de Sitter solution is always unstable. {For the AdS case, Solution 6 does not fulfill the 
Breitenlohner-Freedman bound for generic values of ($\alpha,\beta$).}


\subsection{Type IIB with O5/D5}

Let us now consider type IIB compactification on a (twisted) torus in the presence of an O5 plane, whose configuration is
\begin{align}
\text{O5}: 
\qquad 
\underbrace{\times|\times\times\times\times\times}_{6\mathrm{D}} \ 
\underbrace{- - - - }_{4\mathrm{D}}
\ .
\end{align}
This setting defines the following orientifold involution
\begin{align}
\sigma_{\text{O5}}:\ \ 
y^m \quad \mapsto \quad -y^m &, \ \ m=1,\, 2,\, 3,\, 4\, .
\end{align}

\subsubsection*{Fluxes and moduli}
Because the world-volume of the O5/D5 branes extends along the 6 external dimensions, the SL($4,\bbR$) symmetry arising from the compactification remains unbroken under the $\sigma_{\text{O}5}$ involution. This implies that, in addition to the universal moduli, the scalar fields arising from the internal components of the supergravity fields are encoded in the matrix $\cM_{mn}$, which is a representative of the $\SL(4,\bbR)/\SO(4)$ coset.

\begin{table}[!t]
\centering
\begin{tabular}{|c|c|c|c|c|}
\hline
IIB Flux type & Flux parameters & $\sigma_{\text{KKO5}}$ & no extra $\bbZ_2$ & $\Theta$ components
\\
\hline\hline
$F_{(1)}$		& $F_m\ =\ f_m$		& $+$ 	& $+$ 	& $\zeta_{m}\ =\ f_m$
\\\hline\hline
$H_{(3)} $	& $H_{mnp}\ =\ \epsilon _{mnpq}\ h^q$		& $+$ 	& $+$ 	& $f_{mnp} \ =\ \epsilon _{mnpq}\ h^q$
\\\hline
\end{tabular}
\caption{\textit{
The explicit dictionary between type IIB fluxes consistent with the O5 involution
and deformation parameters of $\cN = (1, 1)$ supergravity in six dimensions.
}}
\label{tab:O5-dict}
\end{table}

Having no additional $\bbZ_2$ parity factors, the set of fluxes that are consistent with the O5 involution consists of
\begin{align}
\begin{array}{rrrrrr}
F_m \ \ \equiv \ \ f_m &,\qquad& H_{mnp} \ \ \equiv \ \ \epsilon _{mnpq}\ h^q&,\qquad& \text{no }F_{(3)}\text{ flux}&,\qquad
\\[5pt]
&&\text{no }F_{(5)}\text{ flux}&,\qquad& \text{no }\omega\text{ flux}&.\qquad
\end{array}
\end{align}
More details are given in Table \ref{tab:O5-dict}.

In total, we have a set of 8 fluxes ($4+4 $) and 11 scalar fields ($1+1+(15-6) $). Additionally, because of the presence of an $\cN=(2,2)$ tadpole due to the presence of D5/O5 sources, the term $V_{\text{D}5/\text{O}5}$ non-trivially contributes to the scalar potential, as the tension can be identified by $T_5\ \equiv \ N_{\mathrm{D}5}\,-\,N_{\mathrm{O}5}=f_m h^m$.

\subsubsection*{Scalar sector and fluxes/embedding tensor dictionary}

Now we will study the mapping between the scalar fields of the compactification given by the coset $\SL(4,\bbR)/\SO(4)$ plus the universal moduli $(\rho,\tau)$ and the scalar fields of the gauged supergravity given by the coset \eqref{eq:sugra-coset}.

\begin{table}[!t]
\centering
\begin{tabular}{|c|c|c|c|}
\hline
IIB fields & $\sigma_{\text{O5}}$ & $\Omega_p$ & \# physical dof's
\\
\hline\hline
$e^m{}_n$	& $+$ 	& $+$ 	& $16-6=10$
\\\hline
$B_{mn} $ 					& $+$ 	& $-$ 	& $-$
\\\hline
$\Phi $ 					& $+$ 	& $+$ 	& $1$
\\\hline\hline
$C_{(0)} $ 					& $+$ 	& $-$ 	& $-$
\\\hline
$C_{(2)} $ 					& $+$ 	& $+$ 	& $6$
\\\hline
$C_{(4)} $ 					& $+$ 	& $-$ 	& $-$
\\\hline
\end{tabular}
\caption{\textit{
Counting of the total 17 propagating scalar dof's allowed by O5-planes in type IIB compactifications down to six dimensions. This is exactly the dimension of the supergravity coset given in \eqref{eq:sugra-coset}, once we subtract from $e^m{}_ n$ the 6 unphysical directions corresponding to the compact $\SO(4)$ generators.
}}
\label{tab:O5-scalars}
\end{table}

The set of scalar fields that are even under the above orientifold involution  is presented in Table \ref{tab:O5-scalars}. The relation between the scalar dof's of gauged supergravity and the ones obtained from compactification is
\begin{align}
\left\{
\begin{array}{rcll}
\Lambda 
&=&
\rho^{4/5}\tau^{6/5}
& ,
\\[5pt]
\Sigma 
&=& 
\rho^{3/10}\tau^{-4/5}
&,
\end{array}
\right.
\qquad
\cH_{MN}
=
\left(
\begin{array}{c|c}
\Lambda \ \cM_{mn}	& 0
\\\hline
0 & \Lambda^{-1} \ \cM^{mn}
\end{array}
\right)
\ .
\end{align}

\begin{table}[t]
\centering
\begin{tabular}{|rc|rcc|}
\hline
\multicolumn{2}{|c|}{$\cN=(2,2)$ QC/tadpoles} & \multicolumn{3}{c|}{Sources}
\\
\hline\hline
$f_m h^m \stackrel{!}{=}0$		& (source)	& D5/O5: 	& $\stackrel{m}{-} \ \stackrel{n}{-} \ \stackrel{p}{-} \ \stackrel{q}{-}$ 	& $(\times 1)$
\\\hline
\end{tabular}
\caption{\textit{
Non-vanishing extra QC \eqref{eq:QC-extra} and their higher-dimensional origin for the flux compactification of type IIB with O5/D5 given in Table \ref{tab:O5-dict}, where BI stands for Bianchi identities and $\tilde \D \equiv \D+\omega\ \wedge  $. A description of the QC as restrictions for the existence of additional local sources is given.}}
\label{tab:O5-tadpoles}
\end{table}

Regarding the fluxes and the consistent deformation parameters of the supergravity theory, the embedding tensor/fluxes dictionary is given in Table \ref{tab:O5-dict}.
Particularly, if we study the QC \eqref{eq:QC} by restricting ourselves to the flux configuration of the table, we find that all of them are straightforwardly satisfied.
On the other hand, the QC \eqref{eq:QC-extra} do not vanish. This implies that the possible critical points are not solutions of the maximal theory. In particular, the presence of an $\cN=(2,2)$ tadpole, which is shown in Table \ref{tab:O5-tadpoles}, precisely justifies the existence of the spacetime filling O5/D5 source that enters the compactification.

\subsubsection*{Critical points}

\begin{table}[!t]
\centering
\begin{tabular}{|c|c|c|c|}
\hline
Sol \# & $f_m$ & $h^m$  & $m^2$
\\\hline\hline
1 	&	$\alpha^m$ 	& $-\alpha^m$ & 
$0_{(\times 13)}$, $|\alpha^m|^2_{(\times 3)}$, $4|\alpha^m|^2_{(\times 1)}$
\\\hline
\end{tabular}
\caption{\textit{
Critical points of the scalar potential induced by the compactification of IIB with O5/D5. In this case, we obtain a unique 4-parameter family of solutions, for which a Mkw vacuum is found.
}}
\label{tab:O5-solutions}
\end{table}

An exhaustive search of the critical points of the scalar potential obtained from the compactification of type IIB with O5/D5 sources has been done. The result consists of a unique 4-parameter family of solutions. The scalar potential evaluated at the critical points vanishes, thus having a Mkw vacuum. Further details on the values of the embedding tensor are given in Table \ref{tab:O5-solutions}.

\subsection{Type IIB with O7/D7}

We study the effective theory arising from the compactification of type IIB theory on a twisted torus in the presence of an O7/D7 source. The source is extended along the following directions:
\begin{align}
\text{O7}: 
\qquad 
\underbrace{\times|\times\times\times\times\times}_{6\mathrm{D}} \ 
\underbrace{\times \times - -}_{4\mathrm{D}}
\ .
\end{align}
This configuration defines the following orientifold involution for the internal coordinates:
\begin{align}
\sigma_{\text{O7}}:\ 
\left\{
\begin{array}{cccll}
y^a &\mapsto & y^a &, \ \ a=1,\, 2\, & ,
\\
y^i &\mapsto & -y^i &, \ \ i=3,\, 4\, & .
\end{array}
\right.
\end{align}

\subsubsection*{Fluxes and moduli}

As a consequence, the $\SL(4,\bbR)$ covariance of the internal manifold is broken down to $\SL(2,\bbR)_L\times \SL(2,\bbR)_R$ by the involution $\sigma_{\text{O}7}$. This  implies that, in addition to the universal moduli, the scalar matrix $\cM_{mn}\in\SL(4,\bbR)/\SO(4)$ is parameterized as
\begin{align}
\cM_{mn}
=
\left(\begin{array}{c|c}
\sigma^2 M_{ab} &
\\
\hline
& \sigma^{-2} \tilde M_{ij}
\end{array}
\right)
\, ,
\label{eq:O7-SL4-matrix}
\end{align}
where $\sigma$ is the modulus describing the relative squeezing between the $ab$ \& the $ij$ cycles, while $M_{ab}$ and $\tilde M_{ij}$ parameterize the cosets $\SL(2,\bbR)_L/\SO(2)$ and $\SL(2,\bbR)_R/\SO(2)$, respectively. Explicit parameterizations can be found, \emph{e.g.}, in \cite{Bergshoeff:2003ri,deRoo:2011fa}.

\begin{table}[!t]
\centering
\begin{tabular}{|c|c|c|c|c|}
\hline
IIB Flux type & Flux parameters & $\sigma_{\text{O7}}$ & $(-1)^{F_L}\Omega_p$ & $\Theta$ components
\\
\hline\hline
\multirow{4}{*}{$\omega$}	& $\omega_{ij}{}^a\ = \ \theta^a\epsilon _{ij}$		& $+$ 	& $+$ 	& $\zeta_{a} \ =\ -\epsilon _{ab}\ \theta^b$
\\\cline{2-5}
& \begin{tabular}{c} $\omega_{ai}{}^j \ =\ (\kappa_a)_i{}^j+\tfrac12 \eta_a \delta_i{}^j$\\ 
$\omega_{ab}{}^c \ =\ -2\eta_{[a}\delta_{b]}{}^c$	  \end{tabular}	& $+$ 	& $+$ 	& 
\begin{tabular}{c}	$f_{ai\bar j} \ =\ (\kappa_a)_i{}^j$ \\
$\xi_{a} \ =\ -\eta_a$ \\
 {$f_{ab\bar c}=\frac 12 \epsilon_{ab}\eta^c$} \end{tabular} 
\\ \hline
$H_{(3)}$		& $H_{abi} \ = \ \epsilon _{ab} \ h_i $		& $-$ 	& $-$ 	& $f_{ab\bar k} \ =\ h^k\, \epsilon _{ab}$
\\\hline\hline
$F_{(1)}$		& $F_a\ =\ f_a$		& $+$ 	& $+$ 	& $f_{ajk}\ = \ f_a\, \epsilon _{jk}$
\\\hline
$F_{(3)}$		& $F_{abi} \ =\ \epsilon _{ab} \ f_i$		& $-$ 	& $-$ 	& $f_{abi}\ = \ f_i\,\epsilon _{ab}$
\\\hline
\end{tabular}
\caption{\textit{
The explicit dictionary between type IIB fluxes consistent with the O7 involution
and deformation parameters of $\cN = (1, 1)$ supergravity in six dimensions. 
{Raising ${\rm SL}(2,\bbR)$ indices has been done via $h^i=\epsilon^{ij}h_k$ and 
$\eta^a=\epsilon^{ab}\eta_b$.} 
}}
\label{tab:O7-dict}
\end{table}

Regarding the fluxes, in addition to the $\sigma_{\text{O}7}$ involution, an additional $\bbZ_2$ parity given by $(-1)^{F_L}\Omega_p$ has to be considered. Then the set of fluxes that are even under the combination of both parities consists of
\begin{align}
\begin{array}{rcllrcllrcll}
\omega_{ij}{}^a&\equiv&\theta^a\epsilon _{ij}&,\quad& \omega_{ai}{}^j&\equiv&(\kappa_a)_i{}^j+\tfrac12 \eta_a \delta_i{}^j&,\quad&
\omega_{ab}{}^c&\equiv&-2\eta_{[a}\delta_{b]}{}^c&,
\\[5pt]
H_{abi}&\equiv&\epsilon _{ab} \ h_i&,\quad& 
F_{a}&\equiv&f_a&,\quad&
F_{abi}&\equiv&\epsilon _{ab} \ f_i&, 
\end{array}
\end{align} 
{with $(\kappa_a)_i{}^i=0$.}
Further details on the components and parity of each field are given in Table \ref{tab:O7-dict}.

In summary we have a set of 16 fluxes ($2+6+2+2+2+2$) and 7 scalar fields ($1+1+1+(3-1)+(3-1) $). As for the scalar potential, the term $V_{\text{O}7/\text{D}7}$ becomes non-trivial, as the tension is identified by
\begin{align}
T_7\ \equiv \ N_{\mathrm{D}7}\,-\,4N_{\mathrm{O}7}=f_a\theta^a
\ ,
\end{align}
where $f_a$ and $\theta_a$ parameterize the $F_{(1)}$ and the metric fluxes, respectively (\emph{c.f.} Table \ref{tab:O7-dict}). As we will see in the following paragraphs, this is a consequence of an $\cN=(2,2)$ tadpole induced by the O7 plane and possible parallel D7 branes.

\subsubsection*{Scalar sector and fluxes/embedding tensor dictionary}

Now we are going to establish the functional relation between the set of scalar fields obtained from the compactification, which are given by the universal sector plus the matrix $\cM_{mn}$ in \eqref{eq:O7-SL4-matrix}, and the set of scalar fields of the gauged supergravity.

\begin{table}[!t]
\centering
\begin{tabular}{|c|c|c|c|}
\hline
IIB fields & $\sigma_{\text{O7}}$ & $(-1)^{F_L}\Omega_p$ & \# physical dof's
\\
\hline\hline
$e^a{}_b \oplus e^i{}_j$	& $+$ 	& $+$ 	& $4+4-(1+1)=6$
\\\hline
$B_{ai} $ 					& $-$ 	& $-$ 	& $4$
\\\hline
$\Phi $ 					& $+$ 	& $+$ 	& $1$
\\\hline\hline
$C_{(0)} $ 					& $+$ 	& $+$ 	& $1$
\\\hline
$C_{ai} $ 					& $-$ 	& $-$ 	& $4$
\\\hline
$C_{abij} $ 				& $+$ 	& $+$ 	& $1$
\\\hline
\end{tabular}
\caption{\textit{
Counting of the total 17 propagating scalar dof's allowed by O7-planes in type IIB compactifications down to six dimensions. This is exactly the dimension of the supergravity coset given in \eqref{eq:sugra-coset}, once we subtract from $e^a{}_ b$ and $e^i{}_ j$ the 2 unphysical directions associated to the $\SO(2)\times \SO(2)$ generators.
}}
\label{tab:O7-scalars}
\end{table}

The scalar fields that survive both $(-1)^{F_L}\Omega_p$ and $\sigma_{\text{O}7}$ projections are presented in Table \ref{tab:O7-scalars}. The functional relation between the scalar fields of gauged supergravity and the ones obtained from compactification reads
\begin{align}
\left\{
\begin{array}{rcll}
\Lambda 
&=&
\tau^2
& ,
\\[5pt]
\Gamma 
&=& 
\rho
& ,
\\[5pt]
\Sigma 
&=& 
\sigma
&,
\end{array}
\right.
\qquad
\cH_{MN}
=
\left(
\begin{array}{c|c}
	\begin{array}{c|c}
	\Lambda \Gamma M_{ab} & 0 
	\\\hline
	0 & \Lambda \Gamma^{-1} M_{ij} 
	\end{array}
	& 0
\\\hline
0 & 
\begin{array}{c|c}
\Lambda^{-1} \Gamma^{-1} M^{ab}  & 0
\\
\hline
0 & \Lambda^{-1} \Gamma M^{ij} 
\end{array}
\end{array}
\right)
\ .
\end{align}
On the other hand, the consistent deformations of the 6-dimensional gauged theory are encoded in the embedding tensor. The explicit parameterization of the fluxes inside the $\SL(2,\bbR)_L\times \SL(2,\bbR)_R\subset \SO(4,4)$ irrep's of the embedding tensor is given in Table \ref{tab:O7-dict}.

\begin{table}[!t]
\centering
\begin{tabular}{|rc|rcc|}
\hline
\multicolumn{2}{|c|}{$\cN=(1,1)$ QC/tadpoles} & \multicolumn{3}{c|}{Sources}
\\
\hline\hline
$\epsilon ^{ab}\left( \eta_a(\kappa_b)_i{}^j-(\kappa_a)_i{}^k(\kappa_b)_k{}^j \right)\stackrel{!}{=}0 $		& (Jacobi)	& KK5/KKO5: 	& $\stackrel{a}{-} \ \stackrel{b}{-} \ \stackrel{i}{-} \ \stackrel{j}{\text{ISO}}$ 	& $(\times 2)$
\\\hline
$\eta_a\theta^b-\frac12 \delta_a^b\eta_c\theta^c \stackrel{!}{=}0 $		& (Jacobi)	& $\widetilde{\text{KK5}}/\widetilde{\text{KKO5}}$: 	& $\stackrel{a}{-} \ \stackrel{b}{\text{ISO}} \ \stackrel{i}{-} \ \stackrel{j}{-}$ 	& $(\times 2)$
\\\hline
$f_a\eta_b\epsilon ^{ab} \stackrel{!}{=}0 $		& (BI $C_{(0)}$)	& $\widetilde{\text{O7}}/\widetilde{\text{D7}}$: 	& $\stackrel{a}{-} \ \stackrel{b}{-} \ \stackrel{i}{\times} \ \stackrel{j}{\times}$ 	& $(\times 1)$
\\\hline\hline\hline
\multicolumn{2}{|c|}{$\cN=(2,2)$ QC/tadpoles} & \multicolumn{3}{c|}{Sources}
\\
\hline\hline
$f_a\theta^a\stackrel{!}{=}0$		& (BI $C_{(0)}$)	& O7/D7: 	& $\stackrel{a}{\times} \ \stackrel{b}{\times} \ \stackrel{i}{-} \ \stackrel{j}{-}$ 	& $(\times 1)$
\\\hline
\end{tabular}

\caption{\textit{
Non-vanishing QC \eqref{eq:QC}, \eqref{eq:QC-extra} and their higher dimensional origin for the flux compactification of type IIB with O7/D7 given in Table \ref{tab:O7-dict}, where BI stands for Bianchi identities and Jacobi refers to the condition \eqref{eq:II-Jacobi}. A description of the QC as restrictions for the existence of additional local sources is given.
}}
\label{tab:O7-tadpoles}
\end{table}

If we study the QC \eqref{eq:QC} by restricting ourselves to such configuration of fluxes, we observe that some of them are not yet satisfied. These conditions, which can be interpreted as restrictions for the presence of additional supersymmetry breaking sources, correspond to the $\cN=(1,1)$ tadpoles written in Table \ref{tab:O7-tadpoles}.
As far as the extra QC \eqref{eq:QC-extra} are concerned, they are not satisfied in general. This implies that, if any, some critical points could genuinely be solutions of the half-maximal theory and not solutions of the maximal one. Similarly, this indicates the existence of an $\cN=(2,2)$ tadpole, which is shown in Table \ref{tab:O7-tadpoles}. The value of the effective tension appearing in the scalar potential is given by $T_7=f_a \theta^a$.

\subsubsection*{Critical points}

\begin{table}[!t]
\centering
\begin{tabular}{|c|c|c|c|c|c|c|c|}
\hline
Sol \# & $f_a$ & $f_i$ & $	h_i$ & $\theta^a$ &  $(\kappa_a)_i{}^j$ & $\eta_a$ & $m^2$
\\\hline\hline
1 	&	$\begin{pmatrix} \alpha \\ 0 \end{pmatrix}$ 	& $0$	& $0$ 		&$\begin{pmatrix} -\alpha\\0\end{pmatrix}$			& $\begin{pmatrix} 0\\\beta \end{pmatrix}\otimes \begin{pmatrix}
0&-1\\1&0 \end{pmatrix}$ & 0 &
\begin{tabular}{c} $0_{(\times 9)}$, $\alpha^2_{(\times 1)}$, $4\alpha^2_{(\times 1)}$, \\
$\beta^2_{(\times 2)}$, $4\beta^2_{(\times 2)}$, 
$(\alpha^2+\beta^2)_{(\times 2)}$ \end{tabular}
\\\hline
2 	&	$0$ 	& $0$	& $0$ 		&$\begin{pmatrix} \alpha\\\beta\end{pmatrix}$			& $\begin{pmatrix} \alpha\\\beta \end{pmatrix}\otimes \begin{pmatrix}
0&1\\0&0 \end{pmatrix}$ & 0 &
$0_{(\times 13)}$, $(\alpha^2+\beta^2)_{(\times 3)}$, $4(\alpha^2+\beta^2)_{(\times 1)}$
\\\hline
3 	&	$0$ 	& $0$	& $0$ 		&$0$	& $\begin{pmatrix} \alpha\\\beta \end{pmatrix}\otimes \begin{pmatrix}
0&-1\\1&0 \end{pmatrix}$ & 0 &
$0_{(\times 11)}$, $(\alpha^2+\beta^2)_{(\times 4)}$, $4(\alpha^2+\beta^2)_{(\times 2)}$
\\\hline
\end{tabular}
\caption{\textit{
Critical points of the scalar potential induced by the compactification of IIB with O7/D7. In this case, we obtain three 2-parameter families of solutions. When evaluated at the critical points, the scalar potential vanishes, thus giving rise to a Mkw vacuum.
}}
\label{tab:O7-solutions}
\end{table}

Let us take a look at the critical points of the scalar potential induced by the fluxes of Table \ref{tab:O7-dict}. 
We find three 2-parameter families of solutions. While the traceless part of the metric flux is turned on for all of them, only one solution  carries 1-form flux. The rest of fluxes vanish. Further details can be found in Table \ref{tab:O7-solutions}.
In Appendix \ref{app:global} the global aspects of the twisted torus compactifications that give rise to each of these vacua are studied.


\subsection{Type IIB with O9/D9}

In this section we study type IIB compactification on a twisted torus with O9/D9 sources. These extended BPS object fills the full 10-dimensional space-time
\begin{align}
\text{O9}: 
\qquad 
\underbrace{\times|\times\times\times\times\times}_{6\mathrm{D}} \ 
\underbrace{\times \times\times\times}_{4\mathrm{D}}
\ ,
\end{align}
and defines a trivial orientifold involution
\begin{align}
\sigma_{\text{O}9}:\ 
\left.
\begin{array}{cccl}
y^m &\mapsto & y^m &, \ \ m=1,\, 2,\, 3,\, 4\, .
\end{array}
\right.
\end{align}

\subsubsection*{Fluxes and moduli}

Since our sources completely fill internal space, the SL($4,\bbR$) covariance emerging from the compactification remains unbroken. Then, in addition to the universal moduli $(\rho,\tau)$, the scalar fields arising from the compactification parameterize a coset, which we denote by $\cM_{mn}\in\SL(4,\bbR)/\SO(4)$.

\begin{table}[!t]
\centering
\begin{tabular}{|c|c|c|c|c|}
\hline
IIB Flux type & Flux parameters & $\sigma_{\text{O9}}$ & $\Omega_p$ & $\Theta$ components
\\
\hline\hline
$\omega$	& $\omega_{mn}{}^p$		& $+$ 	& $+$ 	& $f_{mn\bar p} \ =\ \omega_{mn}{}^p$
\\\hline\hline
$F_3$		& $F_{mnp} \ =\ \epsilon _{mnpq} \ f^q$		& $+$ 	& $+$ 	& $f_{mnp}\ = \ \epsilon _{mnpq} \ f^q$
\\\hline
\end{tabular}
\caption{\textit{
The explicit dictionary between type IIB fluxes consistent with the O9 involution
and deformation parameters of $\cN = (1, 1)$ supergravity in six dimensions.
}}
\label{tab:O9-dict}
\end{table}

The set of fluxes that are consistent with $\Omega_p$ consists of
\begin{align}
\begin{array}{rrrrrr}
\omega_{mn}{}^p&,\qquad&
F_{mnp} \ \equiv \ \epsilon _{mnpq}\ f^q&,\qquad&
\text{no }H_{(3)}\text{ flux}&,
\\
&&
\text{no }F_{(1)}\text{ flux}&,\qquad&
\text{no }F_{(5)}\text{ flux}&.
\end{array}
\end{align}
Further details can be found in Table \ref{tab:O9-dict}.
In summary we have a set of 22 fluxes ($18+4$) and 11 scalar fields ($1+1+(15-6) $). On the other hand, in the scalar potential, the term $V_{\text{O}9/\text{D}9}$ does not contribute, due to the identification $T_9\,\equiv\,N_{\mathrm{D}9}-16N_{\mathrm{O}9}\,=0$\footnote{This is consistent with the standard setup in type I string theory consisting of an O9 plane with 16 parallel D9 branes yielding an anomaly free $\SO(32)$ $\cN=1$ SYM$_{10}$.}.

\subsubsection*{Scalar sector and fluxes/embedding tensor dictionary}

Let us now move to the mapping between scalar fields of the compactification given by the coset $\SL(4,\bbR)/\SO(4)$ and scalar fields of the gauged supergravity given by the coset $\SO(4,4)/\SO(4)\times\SO(4)$.
\begin{table}[!t]
\centering
\begin{tabular}{|c|c|c|c|}
\hline
IIB fields & $\sigma_{\text{O9}}$ & $\Omega_p$ & \# physical dof's
\\
\hline\hline
$e^m{}_n$					& $+$ 	& $+$ 	& $16-6=10$
\\\hline
$B_{mn} $ 					& $+$ 	& $-$ 	& $-$
\\\hline
$\Phi $ 					& $+$ 	& $+$ 	& $1$
\\\hline\hline
$C_{(0)} $ 					& $+$ 	& $-$ 	& $-$
\\\hline
$C_{mn} $ 					& $+$ 	& $+$ 	& $6$
\\\hline
$C_{mnpq} $ 				& $+$ 	& $-$ 	& $-$
\\\hline
\end{tabular}
\caption{\textit{
Counting of the total 17 propagating scalar dof's allowed by O9-planes in type IIB compactifications down to six dimensions. This is exactly the dimension of the supergravity coset given in \eqref{eq:sugra-coset}, once we subtract from $e^m{}_ n$ the 6 unphysical directions corresponding to the compact $\SO(4)$ generators.
}}
\label{tab:O9-scalars}
\end{table}
The set of fields that survive the $\Omega_p$ projection is presented in Table \ref{tab:O9-scalars}. The functional relation between the scalar fields of gauged supergravity and the ones obtained from compactification is
\begin{align}
\left\{
\begin{array}{rcll}
\Lambda 
&=&
\tau^2
& ,
\\[5pt]
\Sigma 
&=& 
\rho^{1/2}
&,
\end{array}
\right.
\qquad
\cH_{MN}
=
\left(
	\begin{array}{c|c}
	\Lambda  \cM_{mn} & 0 
	\\\hline
	0 &  \Lambda^{-1} \cM^{mn} 
	\end{array}
\right)
\ .
\end{align}

The dictionary relating the fluxes and the deformation parameters written as components of the embedding tensor is contained in Table \ref{tab:O9-dict}. These results and the choice $T_9=0$ allows us to unambiguously match the scalar potential \eqref{eq:sugra-potential} with the one obtained from compactification, \eqref{eq:II-potential}.

\begin{table}[!t]
\centering
\begin{tabular}{|rc|rcc|}
\hline
\multicolumn{2}{|c|}{$\cN=(1,1)$ QC/tadpoles} & \multicolumn{3}{c|}{Sources}
\\
\hline\hline
$ \omega_{[mn}{}^r \omega_{p]r}{}^q \stackrel{!}{=}0 $		& (Jacobi)	& KK5/KKO5: 	& $\stackrel{m}{\text{ISO}} \ \stackrel{n}{-} \ \stackrel{p}{-} \ \stackrel{q}{\times}$ 	& $(\times 4)$
\\\hline
\end{tabular}
\caption{\textit{
Non-vanishing QC \eqref{eq:QC} and their higher-dimensional origin for the flux compactification of type IIB with O9/D9 given in Table \ref{tab:O9-dict}, where Jacobi refers to the condition \eqref{eq:II-Jacobi}. A description of the QC as restrictions for the existence of additional local sources is given.
}}
\label{tab:O9-tadpoles}
\end{table}

Upon picking this flux configuration and using the above dictionary, some of the QC associated to the embedding tensor \eqref{eq:QC} are still not automatically satisfied. These conditions, which resemble the restrictions to the presence of additional sources, correspond to the $\cN=(1,1)$ tadpole written in Table \ref{tab:O9-tadpoles}.
Regarding the extra QC \eqref{eq:QC-extra}, because they are straightforwardly zero, we conclude that the hypothetical critical points of the deformed supergravity will also satisfy the equations of motion of the maximal theory.

\subsubsection*{Critical points}

\begin{table}[!t]
\centering\resizebox{\textwidth}{!}{
\begin{tabular}{|c|c|c|c|c|c|c|}
\hline
Sol \# & $f_m$ & $\omega_{mn}{}^1$ & $\omega_{mn}{}^2$ & $\omega_{mn}{}^3$ &  $\omega_{mn}{}^4$ 
& $m^2$
\\\hline\hline			
1 	&	$0$ & $\begin{pmatrix} 0&0&0&0  \\ 0&0&0&0 \\ 0&0&0&\alpha \\ 0&0&-\alpha&0\end{pmatrix}$ 	& $\begin{pmatrix} 0&0&0&0  \\ 0&0&0&0 \\ 0&0&0&\beta \\ 0&0&-\beta&0\end{pmatrix}$	& $\mathbb O_4$ 		&$\begin{pmatrix} 0&0&\alpha&0\\ 0&0&\beta&0 \\ -\alpha&-\beta&0&0 \\ 0&0&0&0 \end{pmatrix}$		
&\begin{tabular}{c} $0_{(\times 11)}$, $(\alpha^2+\beta^2)_{(\times 4)}$,\\
 $4(\alpha^2+\beta^2)_{(\times 2)}$ \end{tabular}			
\\\hline
\end{tabular}}
\caption{\textit{
Some critical points of the scalar potential induced by the compactification of IIB with O9/D9. In this case, 
we show a 2-parameter family of solutions, all of them  corresponding to Minkowski vacua.
}}
\label{tab:O9-solutions}
\end{table}

Before studying the existence of critical points for this configuration, we will prove two more generic results: for type IIB compactifications with spacetime filling O9/D9 sources, (i) all critical points are Minkowski, and (ii) on-shell, $F_{mnp}=0$. 

To prove (i) we just note that the scalar potential can be written as
\begin{align}
V=-\frac12 \partial_\Sigma V
\ .
\end{align}
Then, because a necessary condition for the existence of critical points is precisely $\partial_\Sigma V\stackrel{!}{=}0$, we obtain that
\begin{align}
V_{\text{on-shell}}=0 \ ,
\end{align}
thus concluding that only Minkowski solutions can exist as critical points. \hfill $\Box$

As for the proof of (ii), let us note that the scalar potential can also be written as
\begin{align}
V=-\frac14 f^q f^q -\partial_\Lambda V
\ .
\end{align}
On-shell, the scalar potential reduces to
\begin{align}
V_{\text{on-shell}}
=-\frac14 f^q f^q
\ ,
\end{align}
and, using the result (i), we conclude that $|F_{mnp}|^2=0$, which implies the vanishing of the 3-form flux. \hfill $\Box$

In Table \ref{tab:O9-solutions} we show a single family of critical points. 
Since the 3-form flux is required to vanish on-shell, the solution is sourced only by metric fluxes. 
In addition to the parameterization given in Table \ref{tab:O9-solutions}, one may find other solutions. For instance, the 
following flux configurations also satisfy the conditions for the critical point and QCs: 
\begin{align}
\label{}
\omega_{mn}{}^1&=\omega_{mn}{}^2= \mathbb O_4 \,, \quad 
\omega_{mn}{}^3=\left(
\begin{array}{cccc}
0&0&0&\alpha  \\
0&0&0&\beta  \\
0&0&0&0  \\
-\alpha&-\beta &0 &0  \\
\end{array}
\right)\,, \quad 
\omega_{mn}{}^4=\left(
\begin{array}{cccc}
0&0&-\alpha&0  \\
0&0&-\beta&0  \\
\alpha&\beta&0&0  \\
0&0&0 &0  \\
\end{array}
\right)\,.
\end{align}
Nevertheless, one can show that this flux configuration can be transformed into the 
solution in Table \ref{tab:O9-solutions} by the following change of frame
\begin{align}
\label{O9:framechange}
\omega'_{mn}{}^p=S_m{}^{q}S_n{}^r (S^{-1})_s{}^p 
\omega _{qr}{}^s \,, \qquad 
S_m{}^n= \left(
\begin{array}{cccc}
0&0 &1 &0\\
-\frac{\beta/\alpha}{\sqrt{1+\beta^2/\alpha^2}}& \frac{1}{\sqrt{1+\beta^2/\alpha^2}}&0 &0 \\
\frac{\alpha/\beta}{\sqrt{1+\alpha^2/\beta^2}}&\frac{1}{\sqrt{1+\alpha^2/\beta^2}} & 0& 0\\
0&0 &0 &-1   
\end{array}
\right)\,. 
\end{align}
Other solutions that we found turn out to be equivalent to the solution in Table \ref{tab:O9-solutions}, 
which can be inferred from the degeneracy of mass spectrum. 
Further details concerning the global properties of the solutions, such as the periodic identifications that are necessary to view them as globally well defined compactifications on twisted tori are collected in Appendix \ref{app:global}.

\subsection{M-theory with KKO6/KK6}

Let us finally consider the compactification of 11-dimensional supergravity on twisted tori in the presence of KKO6/KK6 monopoles,
\begin{align}
\text{KKO6}: 
\qquad 
\underbrace{\times|\times\times\times\times\times}_{6\mathrm{D}} \ 
\underbrace{\times \text{ISO} - - - }_{5\mathrm{D}}
\ .
\end{align}
This particular configuration induces the following orientifold involution on the internal coordinates:
\begin{align}
\sigma_{\text{KKO6}}:\ 
\left\{
\begin{array}{cccl}
y^0 &\mapsto & y^0 &, 
\\
y^m &\mapsto & -y^m &, \ \ m=1,\, 2,\, 3,\, 4\, .
\end{array}
\right.
\end{align}
We will assume the presence of an isometry direction along one of the $x^m$ directions. Generically, this setting could be related to the cases of type IIA with KKO5/KK5 and O6/D6 when we turn off the Romans' mass $F_{(0)}=0$, by performing a compactification on a circle along the directions $y^0$ and $y^i=\text{ISO}$, respectively. However, as we will see, the most general flux configuration is still inequivalent, as some of the 11-dimensional fluxes that we are going to consider turn out to be lacking a geometric interpretation in perturbative type IIA.

\subsubsection*{Fluxes and moduli}

The presence of the KKO6 source effectively breaks the SL($5,\bbR$) covariance arising from the dimensional reduction down to $\bbR^+\times\SL(4,\bbR)$. Accordingly, the index $\hat m$ of the fundamental representation of $\SL(5,\bbR)$ introduced in Section \ref{sec:11d-compactification} splits as $\hat m =(0,m)$, with $m=1,\cdots,4$ being an index of the $\mathbf{4}$ of $\SL(4,\bbR)$. Consequently, the non-universal sector of scalar fields that arise from the compactification of the 11-dimensional theory, parameterize the matrix $\hat \cM_{\hat m\hat n}\in \SL(5,\bbR)/\SO(5)$ as follows:
\begin{align}
\hat \cM_{\hat m\hat n}
=
\left(
\begin{array}{c|c}
\sigma^4 & 0
\\\hline
0 & \sigma^{-1} \cM_{mn}
\end{array}
\right)
\ ,
\end{align}
where $\sigma$ is a scalar field and $\cM_{mn}\in \SL(4,\bbR)/\SO(4)$.

\begin{table}[!t]
\centering
\begin{tabular}{|c|c|c|c|}
\hline
11D Flux type & Flux parameters & $\sigma_{\text{KKO6}}$ &  $\Theta$ components
\\\hline\hline
$G_{(4)} $		& $G_{mnpq}=g_4\ \epsilon _{mnpq}$		& $+$ 	& $\zeta_0=g_4$
\\\hline
\multirow{2}{*}{$\omega$}		& $\omega_{mn}{}^0=\theta_{mn}$		& $+$ 	& $\zeta_A=\frac12 \theta_{mn}[G_A]^{mn}$
\\\cline{2-4}
& $\omega_{0m}{}^n=\kappa_m{}^n$		& $+$ 	& $f_{0AB}={-}\kappa_m{}^n[\bar G_A]_{np}[G_B]^{pm}$
\\\hline
\end{tabular}
\caption{\textit{
The explicit dictionary between M-theory fluxes consistent with the KKO6 involution
and deformation parameters of $\cN = (1, 1)$ supergravity in six dimensions.
}}
\label{tab:KKO6-dict}
\end{table}

The presence of KKO6/KK6 monopoles does not introduce any additional $\bbZ_2$ parity \cite{Bergshoeff:2001pv}. Hence, the set of fluxes that are consistent with the above involution $\sigma_{\text{KKO}6}$ is given by
\begin{align}
\begin{array}{rcllrcllrcll}
\omega_{mn}{}^0&\equiv& \theta_{mn}&,\qquad&
\omega_{0m}{}^n&\equiv&\kappa_m{}^n&,\qquad&
G_{mnpq}&\equiv & g_4\ \epsilon _{mnpq}&,
\end{array}
\end{align}
{where $\theta_{mn}=\theta_{[mn]}$ and $\kappa_m{}^m=0$.}
Further details of the parameterization are collected in Table~\ref{tab:KKO6-dict}.
In summary, we have a set of 22 fluxes ($1+6+15$) and 11 scalar fields ($1+1+(15-6) $). In the scalar potential, no extra term $V_{\text{KKO}6/\text{KK}6}$ needs to be included, since KK monopoles are directly sourced by the metric and hence their contribution to the effective potential directly comes from the 11-dimensional Einstein-Hilbert term.

\subsubsection*{Scalar sector and fluxes/embedding tensor dictionary}

Let us firstly study the dictionary between the scalar fields of the compactification, which are given by the coset $\bbR^+\times\bbR^+\times \SL(4,\bbR)/\SO(4)$ and the scalar fields of the gauged supergravity parameterizing the coset $\bbR^+ \times \SO(4,4)/\SO(4)\times\SO(4)$. 

\begin{table}[!t]
\centering
\begin{tabular}{|c|c|c|c|}
\hline
11D fields & $\sigma_{\text{KKO6}}$  & \# physical dof's
\\
\hline\hline
$e^0{}_0\ \oplus\ e^m{}_n$	& $+$ 	& $1+16-6=11$
\\\hline
$A_{0mn} $ 					& $+$ 	& $6$
\\\hline
\end{tabular}
\caption{\textit{
Counting of the total 17 propagating scalar dof's allowed by KKO6 sources in M-theory compactifications down to six dimensions. This is exactly the dimension of the supergravity coset given in \eqref{eq:sugra-coset}, once we subtract from $e^m{}_ n$ the 6 unphysical directions corresponding to the compact $\SO(4)$ generators.
}}
\label{tab:KKO6-scalars}
\end{table}

The set of fields that are even under the above involution is presented in Table \ref{tab:KKO6-scalars}. As in the case of type IIA with KKO5, the mapping relating the scalar fields of each formulation is spinorial and therefore it may be established by making use of the isomorphism $\mathfrak{sl}(4,\bbR)\cong \mathfrak{so}(3,3)$. To do so, we use the same splitting of the $\SO(4,4)$ light-cone coordinates as the one done in \eqref{eq:SO44-to-SO33}. Then, the supergravity scalar fields are parameterized as
\begin{align}
\left\{
\begin{array}{rcll}
\Lambda 
&=&
\rho^{3}\sigma^2
& ,
\\[5pt]
\Sigma 
&=& 
\rho^{-3/8} \sigma 
&,
\end{array}
\right.
\qquad
\cH_{MN}
=&
\left(
\begin{array}{c|c}
	\begin{array}{c|c}
	\Lambda  & 0 
	\\\hline
	0 &{ \cM_{ij} }
	\end{array}
	& 0
\\\hline
0 & 
\begin{array}{c|c}
\Lambda^{-1}  & 0
\\
\hline
0 & {\cM^{\bar i\bar j}}
\end{array}
\end{array}
\right)
\ ,
\end{align}
where {$\cM_{ij}$ and $\cM^{\bar i\bar j}$ are the components of $\cM_{AB}$  given by \eqref{eq:SO33-scalars}}. Similarly, the vielbein $\cV_A{}{}^{\underline{I} \underline{\hat J} }$ that squares to $\cM_{AB}$ is  the one given by \eqref{eq:SO33-vielbein}.
\begin{table}[!t]
\centering
\begin{tabular}{|rc|rcc|}
\hline
\multicolumn{2}{|c|}{$\cN=(1,1)$ QC/tadpoles} & \multicolumn{3}{c|}{Sources}
\\
\hline\hline
$ \theta^{m[n}\kappa_m{}^{p]} \stackrel{!}{=}0 $		& (Jacobi)	& ?? &   (no long weights in the $\mathbf{6}$) &
\\\hline
$ \theta^{m(n}\kappa_m{}^{p)} \stackrel{!}{=}0 $		& (Jacobi)	& $\widetilde{\mathrm{KK6}}/\widetilde{\mathrm{KKO6}}$: 	& $\stackrel{0}{\times} \ \stackrel{m}{\text{ISO}} \ \stackrel{n}{-} \ \stackrel{p}{-} \ \stackrel{q}{-}$ 	& $(\times 4)$
\\\hline
\end{tabular}
\caption{\textit{
Non-vanishing QC \eqref{eq:QC} and their higher dimensional origin for M-theory with KKO6/KK6 given in Table \ref{tab:KKO6-dict}. Note that the source interpretation for the QC in the first line is not clear. This is generically the case whenever the corresponding QC transforms in irrep's which do not contain long weights \cite{Bergshoeff:2011ee,Lombardo:2017yme}. These objects might though correspond to non-trivial bound states of elementary branes. {In the table, we have denoted $\theta^{mn} \equiv \frac 12 \epsilon^{mnpq}\theta_{pq}$.}}}
\label{tab:KKO6-tadpoles}
\end{table}
Regarding the internal components of the fields and the compatible deformations of the theory, a detailed dictionary between the consistent fluxes and the embedding tensor components is spelled out in Table \ref{tab:KKO6-dict}.

Let us consider the QC \eqref{eq:QC} for the set of fluxes of Table \ref{tab:KKO6-dict}. We find that some conditions are not automatically satisfied. Such equations forbid the presence of additional sources that will not preserve the 16 supercharges of the theory. In particular, these expressions are written in Table \ref{tab:KKO6-tadpoles} and correspond to the tadpoles of the  $\cN=(1,1)$ theory. These tadpoles are precisely the long weights of the $\mathbf{10}$ (4 states), which prohibit the presence of KKO6/KK6 monopoles.

Finally, let us consider the extra QC \eqref{eq:QC-extra}, which determines whether a deformation of the half-maximal theory is also consistent in the maximal case. Plugging the non-vanishing components of the embedding tensor we find that they are all satisfied. This means that such solutions will also be solutions of the maximal theory.

\subsubsection*{Type IIA/M-theory duality}

As we have mentioned above, upon doing a compactification on a circle either along the direction of the KKO6 world-volume $y^0$ ($S_0^1$) or the isometry direction, say $y^1$, ($S^1_1$), a mapping between M-theory and type IIA configurations can be established.

Let us firstly note that, depending on which compactification circle we pick, $S^1_0$ or $S^1_1$, the KKO6-plane induces two types of local sources:
\begin{align}
\text{KKO6}: \quad
\stackrel{0}{\times} \ 
\stackrel{1}{\text{ISO}} \ 
\stackrel{i}{-} \ 
\stackrel{j}{-} \ 
\stackrel{k}{-} 
\qquad 
\Longrightarrow
\qquad
\left\{
\begin{array}{crr}
\stackrel{S^1_0}{\longrightarrow} & \text{KKO5:} & \stackrel{1}{\text{ISO}} \ \stackrel{i}{-} \ \stackrel{j}{-} \ \stackrel{k}{-} 
\\[5pt]
\stackrel{S^1_1}{\longrightarrow} & \text{O6:} & \stackrel{0}{\times} \ \stackrel{i}{-} \ \stackrel{j}{-} \ \stackrel{k}{-} 
\end{array}
\right.
\ .
\end{align}

Secondly, using the Kaluza-Klein Ansatz for the dimensional reduction of M-theory on a circle, we can easily read off the resulting 10-dimensional fields. For example, for the compactification along the direction $y^0$, the 11-dimensional fields turn on the following type IIA fluxes:
\begin{align}
\left.
\begin{array}{rclcccrclcccrclc}
G_{1ijk}&\rightarrow& F_{1ijk}
&,&&
\omega_{0i}{}^1&\rightarrow& \text{non-geom.}
&,&&
\omega_{ij}{}^0&\rightarrow& F_{ij}
&,
\\[5pt]
\omega_{01}{}^i&\rightarrow& \text{non-geom.}
&,&&
\omega_{0i}{}^j&\rightarrow& \text{non-geom.}
&,&&
\omega_{1i}{}^0&\rightarrow& F_{1i}
&.
\end{array}
\right.
\end{align}
Therefore, following the parameterizations of Tables \ref{tab:KKO6-dict} and \ref{tab:KKO5-dict}, the fluxes are related as follows:
\begin{align}
g_4 =f_4 
\ ,
\qquad
\theta_{mn}=F_{mn}
\ ,
\qquad
\kappa_{m}{}^n=\text{non-geom.}
\ .
\end{align}
We observe that some 11-dimensional metric fluxes have no geometric analogue in type IIA, as they would correspond to strong coupling effects within the KKO5 truncation. 

Similarly, for the compactification along the isometry direction $y^1$, we obtain the following relations:
\begin{align}
\left.
\begin{array}{rclccrclccrclc}
G_{1ijk}&\rightarrow& H_{ijk}
&,& &
\omega_{0i}{}^1&\rightarrow& F_{0i}
&,& &
\omega_{ij}{}^0&\rightarrow& \omega_{ij}{}^0
&,
\\[5pt]
\omega_{01}{}^i&\rightarrow& \text{non-geom.}
&,&&
\omega_{0i}{}^j&\rightarrow& \omega_{0i}{}^j
&,&&
\omega_{1i}{}^0&\rightarrow& \text{non-geom.}
&.
\end{array}
\right.
\end{align}
In this case, the parameterizations of Tables \ref{tab:KKO6-dict} and \ref{tab:O6-dict} are related as:
\begin{align}
g_4 =h 
\ , \qquad
\theta_{mn} = \left(\begin{array}{c|c}
0&\text{non-geom.}\\\hline
\text{non-geom.}& \theta_{ij}
\end{array}\right)
\ ,
\qquad
\kappa_{m}{}^n
=
\left(\begin{array}{c|c}
-\kappa_0&\text{non-geom.}\\\hline
f_i& \kappa_i{}^j\oplus \kappa_0
\end{array}\right)
\ .
\end{align}
As in the previous case, some metric fluxes cannot be mapped to any 10-dimensional (perturbative) flux, thus making this compactification genuinely 11-dimensional.

\subsubsection*{Critical points}

\begin{table}[!t]
\centering
\begin{tabular}{|c|c|c|c|c|}
\hline
Sol \# & $g_4$ & $\theta_{mn}$ & $\kappa_m{}^n$  & $m^2$
\\\hline\hline
1 	&	$0$ &  $\begin{pmatrix} 0&\beta+\frac{\alpha^2}{\beta}&0&0 \\ -\beta-\frac{\alpha^2}{\beta}&0&0&0 \\ 0&0&0&0 \\ 0&0&0&0 \end{pmatrix}$		& $\begin{pmatrix} -\alpha& {\frac{\alpha^2}{\beta}}&0&0 \\ {-\beta}&\alpha&0&0 \\ 0&0&0&0 \\ 0&0&0&0 \end{pmatrix}$
& \begin{tabular}{c}$0_{(\times 13)}$, $\beta^{-2}(\alpha^2+\beta^2)_{(\times 3)}$, 
\\ $4\beta^{-2}(\alpha^2+\beta^2)_{(\times 3)}$\end{tabular}
\\\hline
2 	&	$0$ &  $\begin{pmatrix} 0&\beta+\frac{\alpha^2}{\beta}&0&0 \\ -\beta-\frac{\alpha^2}{\beta}&0&0&0 \\ 0&0&0&0 \\ 0&0&0&0 \end{pmatrix}$		& $\begin{pmatrix} 0&0&0&0 \\ 0&0&0&0 \\ 0&0&\alpha&{\frac{\alpha^2}{\beta} }\\ 0&0&{-\beta}&{-\alpha} \end{pmatrix}$
&\begin{tabular}{c}$0_{(\times 13)}$, $\beta^{-2}(\alpha^2+\beta^2)_{(\times 3)}$, \\
$4\beta^{-2}(\alpha^2+\beta^2)_{(\times 3)}$ \end{tabular}
\\\hline
3 	&	$0$ &  $\mathbb O_4$ & $\begin{pmatrix} 0&\alpha&0&0 \\ -\alpha&0&0&0 \\ 0&0&0&\beta \\ 0&0&-\beta&0 \end{pmatrix}$		&
\begin{tabular}{c}
$0_{(\times 9)}$, $4\alpha^2_{(\times 2)}$, $4\beta^{2}_{(\times 2)}$, \\
$(\alpha+\beta)^2_{(\times 2)}$, $(\alpha-\beta)^2_{(\times 2)}$ \end{tabular}
\\\hline
\end{tabular}
\caption{\textit{
Critical points of the scalar potential induced by the compactification of M-theory with KKO6/KK6. In this case, we show three families of solutions depending on two parameters $(\alpha,\beta)$, all of them  corresponding to Minkowski vacua. 
}}
\label{tab:KKO6-solutions}
\end{table}

The set of critical points of the scalar potential induced by the fluxes of Table \ref{tab:KKO6-dict} is given in Table \ref{tab:KKO6-solutions}. 
We find that the 4-form flux vanishes for all families of solutions that have been found, so the vacua are induced by purely metric flux compactification. In addition, every family only contains Minkowski extrema.
For the solution 2, the Jacobi identity $\theta^{m(n}\kappa_m{}^{p)}=0$ fails to be satisfied, 
this implying the existence of a KKO6 plane\footnote{The corresponding effective tension turns out to be negative, the associated background geometry being Atiyah-Hitchin space \cite{Atiyah:1985dv}.}. For the rest of solutions, the corresponding internal manifold is discussed in
Appendix \ref{app:global}.

\section{Conclusions}
\label{sec:conclusions}

We have studied various aspects of type-II and M-theory compactifications down to six dimensions that explicitly break half of the supersymmetry through the presence of spacetime filling orientifold planes. 
The reduced $D=6$ theory admits the gauged $\cN=(1,1)$ supergravity description.
Note that such 6D theory is always \emph{nonchiral}, regardless of the chirality property of the progenitor theory in ten/eleven dimensions. This is by construction imposed by our truncation procedure that realizes the supersymmetry halving. 
In particular, due to the nonexistence of consistent deformations of $\cN=(2,0)$ theory in six dimensions \cite{Bergshoeff:2007vb}, this enforces the nontriviality of the problem of moduli stabilization when reducing down to 6D, thanks to the presence of nonvanishing background fluxes. 

We have studied various cases obtained by restricting the embedding tensor to components admitting a higher dimensional interpretation within different orientifold compactifications. 
After writing down the corresponding scalar potentials for the wouldbe moduli fields, we have examined the critical points by using the framework of six dimensional gauged supergravities. 
In most of the cases under study, the $D=6$ theories only admit Minkowski vacua. An exceptional case is the massive IIA with KKO5/KK5, for which we possess a rich vacuum structure as displayed in Table \ref{tab:KKO5-solutions-masses}. 
In particular, there exist de Sitter extrema. Note that the corresponding setup goes beyond the conventional framework of \cite{Gibbons:1984kp,Maldacena:2000mw}, in which their existence at a classical level is systematically ruled out. 
However, consistently with the refined no-go argument of \cite{Dasgupta:2019gcd} (see also \cite{Dasgupta:2019rwt}), our de Sitter solutions suffer from tachyonic instabilities. 

All our constructions are based on the presence of spacetime filling orientifold planes and (possibly) parallel positive tension branes of the same type. Within this context, a very natural follow-up question is to wonder what happens to the dynamics of the corresponding compactifications once open string degrees of freedom (\emph{i.e.} brane position moduli and axions obtained by reducing gauge fields on internal cycles) and the corresponding fluxes (\emph{i.e.} non-Abelian open string gaugings) are included as well. Will there appear new structures in this portion of the Landscape? Are there any candidate metastable non-supersymmetric AdS vacua to test swampland conjectures? Can one trust these vacua? We hope to come back to these issues in the future.

\section*{Acknowledgments}

The work of GD is partially supported by the Spanish government grant MINECO-16-FPA2015-63667-P. He also acknowledges support from the Principado de Asturias through the grant FC-GRUPIN-IDI/2018/000174.
The work of JJF-M is supported by Universidad de Murcia-Plan Propio Postdoctoral. 
The work of MN is supported by Grant-in-Aid for Scientific Research (A) from JSPS 17H01091.
GD \& JJFM acknowledge  the financial  support  of  Spanish  Ministerio  de  Econom\'ia  y Competitividad  and  CARM  Fundaci\'on  S\'eneca  under grants FIS2015-28521 and 21257/PI/19.
GD \& MN would also like to thank the Physics Department at the University of Murcia for its warm hospitality while a substantial part of this project was carried out.

\appendix

\section{A $\mathbb{Z}_2$ truncation of maximal gauged supergravity}
\label{app:truncation}

This appendix presents the prescription for obtaining \emph{nonchiral} half-maximal gauged supergravity 
from the maximal gauged supergravity constructed in \cite{Bergshoeff:2007ef}. 
The maximal ($2,2$) gauged supergravity possesses an on-shell $\mathrm{E}_{5(5)}={\rm SO}(5,5)$ U-duality 
symmetry \cite{Tanii:1984zk} and contains 16 vector fields carrying a chiral spinor representation ${\bf 16}_c $
of ${\rm SO}(5,5)$. 

The truncation to ${\cal N}=(1,1) $ half maximal supergravity has been discussed in \cite{Bergshoeff:2007ef}.
Here we repropose a detailed description of this truncation. Apart from practical utilities, 
this was also used for the derivation of the scalar potential presented in (\ref{eq:sugra-potential}). 
We would like to stress that not every  ${\cal N}=(1,1) $  theory is obtained by truncation of ${\cal N}=(2,2) $ theory, 
since the  ${\cal N}=(1,1) $  theory admits a much wider range of possibilities. As a consequence, an extra set of quadratic constraints 
on the embedding tensor appear upon truncations, which we shall discuss in the following.

\subsection{${\rm SO}(5,5)$ branching rules}

Let $\eta_{\mathbb{MN}}$ denote the ${\rm SO}(5,5)$ invariant metric in the light cone basis. 
We split the fundamental indices of ${\rm SO}(5,5)$ as $\mathbb{M}\to (-,\,M,\, +)$, where $M$ is a fundamental ${\rm SO}(4,4)$ index. Correspondingly the metric becomes
\begin{align}
\label{SO55metric}
\eta_{\mathbb{MN}}=\eta^{\mathbb{MN}}=\left(
\begin{array}{c|c|c}
0      &  0 & 1 \\ \hline
 0     &   \eta_{MN} & 0\\ \hline
1 & 0 & 0
\end{array}
\right)\ , \qquad \eta_{MN}= \left(
\begin{array}{cc}
\mathbb   O_4   &  \mathds{1}_4  \\
\mathds{1}_4    & \mathbb  O_4
\end{array}
\right)\ . 
\end{align}
The ${\rm SO}(5,5)$ algebra is characterized by the generators $M_{\mathbb{MN}}$ satisfying 
\begin{align}
\label{N8_SO55}
[M_{\mathbb{MN}}, M_{\mathbb{PQ}}]=2(\eta_{\mathbb{M[P}}M_{\mathbb{Q]N}}-\eta_{\mathbb{N[P}}M_{\mathbb{Q]M}})\ .
\end{align}
Let $\Phi$ denote any ${\rm SO}(5,5)$ field, which transform as 
\begin{align}
\label{N8_SO55rep}
[M_{\mathbb{MN}},\Phi]=-M_{\mathbb{MN}}(\Phi) \ . 
\end{align}
In the following, 
we wish to determine the branching of irrep's of ${\rm SO}(5,5)$  under $\mathbb R^+\times {\rm SO}(4,4)$. 

\subsubsection*{\bf Vector}

In the fundamental representation, the generators can be chosen to be 
$(M_{\mathbb{MN}})_{\mathbb P}{}^{\mathbb Q}=-2\delta ^{\mathbb Q}{}_{[\mathbb M}\eta_{\mathbb{N]P}}$. 
Substituting this into (\ref{N8_SO55rep}), the ${\rm SO}(5,5)$ vector $V_{\mathbb M}$ obeys
$[M_{\mathbb{MN}},V_{\mathbb P}]=-(M_{\mathbb{MN}})_{\mathbb P}{}^{\mathbb Q}V_{\mathbb Q}=-2\eta_{\mathbb{P[M}}V_{\mathbb N]}$. 
Defining the $\mathbb R^+\simeq{\rm SO}(1,1)$ generator by 
\begin{align}
\label{}
D=M_{+-}\ , 
\end{align}
and setting $V_{\mathbb M}=(V_-,  V_M, V_+)$, 
the above relation decomposes into
\begin{align}
\label{}
[D, V_\pm ]=\pm V_\pm\,, \qquad [D, V_M]=0 \ . 
\end{align}
This implies that we can assign $\pm 2$ $\mathbb R^+$ weights\footnote{ 
The weight $\pm 2$ has been chosen for convenience in such a way that 
the spinor representation has $\pm 1$ $\mathbb R^+$ weights. See, (\ref{SO55spinordec88}).}
to the two singlets of  ${\rm SO}(4,4)$ as
\begin{align}
\label{}
{\bf 10}\to{\bf 1}^{(+2)}\oplus{\bf 1}^{(-2)}\oplus{\bf 8}^{(0)}_{v} \ . 
\end{align}

\subsubsection*{Adjoint}

The decomposition of adjoint representation can be read off from (\ref{N8_SO55}). Defining
$ P_M= M_{+M} $ and $K_M=M_{-M}$, 
we obtain
\begin{align}
&~~\quad [D, P_M]= P_M \ , \qquad \qquad \qquad \qquad \qquad  [D, K_M]=-K_M \ , \\
&\quad [P_M, K_N]=-(D\eta_{MN}+M_{MN})\ ,\qquad \ 
[M_{MN}, P_P]=-2\eta_{P[M}P_{N]} \ ,\\
&[M_{MN}, K_P]=-2\eta_{P[M}K_{N]} \ ,  \qquad \qquad \ \
[M_{MN}, M_{PQ}]=2(\eta_{M[P}M_{Q]N} -\eta_{N[P}M_{Q]M})\ , 
\end{align}
while other commutators vanish. 
It follows that 
\begin{align}
\label{}
{\bf 45}\to & {\bf 1}^{(0)}\oplus{\bf 28}^{(0)}\oplus{\bf 8}_v^{(+2)}\oplus{\bf 8}_v^{(-2)}\,. 
\end{align}

\subsubsection*{Spinor}

Let us consider the 32 dimensional ${\rm SO}(5,5)$ spinor representation
\begin{align}
\label{}
(M_{\mathbb{MN}})_{\mathcal A}{}^{\cal B}=\frac 12 (\mathtt\Gamma_{\mathbb{MN}})_{\mathcal A}{}^{\cal{B}}\ , \qquad 
[M_{\mathbb{MN}}, Q_{\mathcal A}]=-\frac 12 (\mathtt\Gamma_{\mathbb{MN}})_{\mathcal A}{}^{\cal B} Q_{\cal B}\ .
\end{align}
where $\mathcal A=1,...,32$ and 
$\mathtt\Gamma_{\mathbb{MN}}=\mathtt\Gamma_{\mathbb{[M}}\, \mathtt\Gamma_{\mathbb{N]}}$. Here, $\mathtt\Gamma_{\mathbb{M}}$ satisfies the 
${\rm SO}(5,5)$ Clifford algebra 
\begin{align}
\label{}
\{\mathtt\Gamma_{\mathbb{M}} , \mathtt\Gamma_{\mathbb{N}} \}=2\eta_{\mathbb{MN}}\,\mathds{1}_{32} \ . 
\end{align}
We employ the following explicit representation 
\begin{align}
\label{SO55gamma}
\mathtt \Gamma_{\underline 1}&= \sigma_{13444}\,, \quad 
\mathtt\Gamma_{\underline{\#}}=i \sigma_{24444}\,,  \quad 
\mathtt \Gamma_{\underline 2}=\sigma_{11212}\,, \quad 
\mathtt \Gamma_{\underline 3}=\sigma_{11131}\,, \quad 
\mathtt \Gamma_{\underline 4}=\sigma_{11114}\,, \notag \\
\mathtt \Gamma_{\underline 5}&=\sigma_{11133}\,, \quad 
\mathtt \Gamma_{\underline 6}=i\sigma_{12444}\,, \quad 
\mathtt \Gamma_{\underline 7}=i\sigma_{11312}\,, \quad 
\mathtt \Gamma_{\underline 8}=i\sigma_{11424}\,, \quad 
\mathtt \Gamma_{\underline 9}=i\sigma_{11432}\,,
\end{align}
where 
$\sigma_{ijklm}=\sigma_i\otimes\sigma_j\otimes\sigma_k\otimes\sigma_l\otimes\sigma_m$
and  
$\sigma_i=(\sigma_1,\sigma_2,\sigma_3, \mathds{1}_2)$. 
The indices with underbar are intended to denote the diagonal ${\rm SO}(5,5)$ invariant metric
$\eta_{\underline{\mathbb{MN}}}={\rm diag}(\mathds{1}_5, -\mathds{1}_5)$. 
The transformation to the indices at hand (\ref{SO55metric}) can be done via 
\begin{align}
\label{}
\mathtt\Gamma_\mp=&\frac {\mathtt\Gamma_{\underline 1}\mp \mathtt\Gamma_\#}{\sqrt 2}\,, \quad 
\mathtt\Gamma_{M}=\frac{\mathtt\Gamma_{\underline M}- \mathtt\Gamma_{\underline{M+4}}}{\sqrt 2}\,, \quad 
\mathtt\Gamma_{M+4}=\frac{\mathtt\Gamma_{\underline M}+ \mathtt\Gamma_{\underline{M+4}}}{\sqrt 2}
\quad (\underline{M}=2,..,5) \,.
\end{align}
The ${\rm SO}(5,5)$ charge conjugation matrix $\mathtt C$ satisfies
\begin{align}
\label{}
\mathtt\Gamma_{\mathbb M}\mathtt C=(\mathtt\Gamma_{\mathbb M}\mathtt C)^T,\qquad  
\mathtt C_{\cal{AB}}=-\mathtt C_{\cal{BA}}\,.
\end{align}
The spinor indices are raised and lowered by $\mathtt C_{\cal{AB}}$ 
and its inverse transpose $\mathtt C^{\cal{AB}}$  as 
$Q^{\cal A}=\mathtt C^{\cal{AB}}Q_{\cal B}$ and 
$Q_{\cal A}=Q^{\cal B}\mathtt C_{\cal{BA}}$. 
Explicitly,  we have 
\begin{align}
\label{}
\mathtt C_{\cal{AB}}=
\left(
\begin{array}{cc}
      & \mathtt C_{A}{}^{B'}  \\
\mathtt C^{A'}{}_{B}    &   
\end{array}
\right) \,, \qquad 
\mathtt C^{\cal{AB}}=
\left(
\begin{array}{cc}
      & \mathtt C^{A}{}_{B'}  \\
\mathtt C_{A'}{}^{B}    &   
\end{array}
\right)\,, \qquad 
\mathtt C_{\cal{AB}}=\mathtt C^{\cal{AB}}=i \sigma_{23344}\,,
\end{align}
with $ \mathtt C_A{}^{C'} \mathtt C_{C'}{}^B=-\delta_A^B $ and 
$ \mathtt C^{A'}{}_C \mathtt C^C{}_{B'}=-\delta ^{A'}_{B'} $, 
 yielding 
\begin{align}
\label{}
 Q_{\mathcal A}= \left(
 \begin{array}{c}
 Q_A    \\
Q^{A'}   
\end{array}
 \right) = \left(
 \begin{array}{c}
 \mathtt C^{B'}{}_A Q_{B'}   \\
 \mathtt C_B{}^{A'}Q^B
\end{array}
 \right)
 \,, \qquad 
  Q^{\cal A}= \left(
 \begin{array}{c}
\mathtt C^A{}_{B'}Q^{B'}   \\
\mathtt C_{A'}{}^BQ_B 
\end{array}
 \right) \,.
\end{align}
In this representation, the Majorana spinor is real
$Q= (\mathtt C\bar Q^T)=\mathtt C(\mathtt \Gamma^{\underline{6789}\#})^TQ^*=Q^* $
and the ${\rm SO}(5,5)$ chiral matrix takes the diagonal form
\begin{align}
\label{}
\mathtt\Gamma_{*}\equiv \mathtt\Gamma_{\underline{1}\cdots \underline{9}\#}
 = {\rm diag}(-\mathds{1}_{16}, +\mathds{1}_{16})
 \end{align}
 Therefore, 
the gamma matrices are chirally decomposed as
\begin{align}
\label{SO55_gamma}
(\mathtt\Gamma_{\mathbb M})_{\cal A}{}^{\cal B}=\left(
\begin{array}{cc}
      &  (\Gamma_{\mathbb M})_{AA'}   \\
 ({\Gamma}_{\mathbb M})^{A'A}   &   
\end{array}
\right)\,, 
\end{align}
so that 
\begin{align}
\label{}
(M_{\mathbb{MN}})_{\cal A}{}^{\cal B}
=\left(
\begin{array}{cc}
 \frac 12 (\Gamma_{\mathbb{MN}})_A{}^B      &    \\
      &   \frac 12 (\Gamma_{\mathbb{MN}})^{A'}{}_{B'}
\end{array}
\right)\,,
\end{align}
where
\begin{align}
\label{}
(\Gamma_{\mathbb{MN}})_A{}^B&=\frac12[ (\Gamma_{\mathbb M})_{AA'}(\Gamma_{\mathbb N})^{A'B}
-(\Gamma_{\mathbb N})_{AA'}(\Gamma_{\mathbb M})^{A'B}]\,, \notag \\ 
(\Gamma_{\mathbb{MN}})^{A'}{}_{B'} &=\frac 12 [({\Gamma}_{\mathbb M})^{A'A}(\Gamma_{\mathbb N})_{AB'}
-({\Gamma}_{\mathbb N})^{A'A}(\Gamma_{\mathbb M})_{AB'}
].
\end{align}
In the hereafter, we focus on $(M_{\mathbb{MN}})^{A'}{}_{B'}$, i.e., 
${\bf 16}_c$.

Let us decompose $(M_{\mathbb{MN}})^{A'}{}_{B'}$
into irrep. of $\mathbb R^+\times {\rm SO}(4,4)$. 
The $ {\rm SO}(4,4)$ gamma matrices $(\mathbf \Gamma_M)^{A'}{}_{B'}$ can be extracted from 
$\mathtt \Gamma_{2- 9}$ as 
\begin{align}
\label{}
\mathtt \Gamma_{M}=\sigma_1\otimes \mathbf \Gamma_{M-1} \quad (M=2,...,9)\,, \qquad 
\{\mathbf \Gamma_M, \mathbf \Gamma_N\}=2\eta_{MN} \mathds{1}_{16} \,. 
\end{align}
The ${\rm SO}(4,4)$ chiral matrix is 
\begin{align}
\label{}
\mathbf \Gamma_9\equiv \mathbf \Gamma_{\underline{1\cdots 8}}={\rm diag}(\mathds{1}_8, -\mathds{1}_8)\,, \qquad 
Q^{A'}=\left(
\begin{array}{c}
Q^{\dot\alpha}   \\
Q_{\alpha}   
\end{array}
\right) \,, \qquad \alpha , \dot \alpha=1,...,8 \,, 
\end{align}
where the dotted indices $\dot\alpha ,\dot\beta,...$ refer to ${\bf 8}_c$ while undotted indices $\alpha ,\beta,...$ refer to ${\bf 8}_s$. 
The ${\rm SO}(4,4)$ charge conjugation matrix $\mathbb C^{A'B'}=\mathtt C_A{}^{B'}\delta^{AA'}$ satisfies
\begin{align}
\label{}
(\mathbf \Gamma_M\mathbb C)=( \mathbf \Gamma_M\mathbb C)^T \,, \qquad \mathbb C=\mathbb C^T \,, \qquad 
\mathbb C=\left(
\begin{array}{cc}
\mathbb C^{\dot \alpha\dot\beta}   &    \\
      &   \mathbb C_{\alpha\beta}  
      \end{array}
\right)  \,,
\end{align}
where $\mathbb C^{\dot \alpha\dot\beta}=-\mathbb C_{\alpha\beta} =\sigma_{344}={\rm diag}(\mathds{1}_4,-\mathds{1}_4)$. 
In the current representation, $(M_{+-})^{A'}{}_{B'}$ reads 
\begin{align}
\label{}
(M_{+-})^{\dot\alpha}{}_{\dot\beta}=-\frac 12\delta^{\dot\alpha}{}_{\dot\beta} \,,\qquad 
(M_{+-})_\alpha{}^\beta=+\frac 12\delta_\alpha{}^\beta\,, \qquad 
\end{align}
leading to
\begin{align}
\label{}
[D, Q^{\dot\alpha}]=+\frac 12 Q^{\dot\alpha}\,, \qquad 
[D, Q_{ \alpha}]=-\frac 12 Q_{\alpha}\,.
\end{align}
One therefore arrives at 
\begin{align}
\label{SO55spinordec88}
{\bf 16}_c \to  & \ {\bf 8}_c^{(+1)}\oplus{\bf 8}_s^{(-1)}\,.
\end{align}

\subsubsection*{Vector-spinor}

We next wish to decompose the ${\rm SO}(5,5)$ vector spinor 
$\theta^{A'}{}_{\mathbb{M}}\in{\bf 144}_c$, which transforms according to 
\begin{align}
\label{}
[M_{\mathbb{MN}}, \theta^{A'}{}_{\mathbb P}]=2\theta^{A'}{}_{[\mathbb{M}}\eta_{\mathbb{N]P}}- (M_{\mathbb{MN}})^{A'}{}_{B'}\theta^{B'}{}_{\mathbb P}\,. 
\end{align}
Here $\theta^{A'}{}_{\mathbb M}$ is a $16\times 10$ matrix subjected to 
\begin{align}
\label{LCmaxSUGRA}
0=(\Gamma_{\mathbb M})_{AB}\theta^{B\mathbb M}=(\Gamma_{\mathbb M})_{AB'}\theta^{B'\mathbb M}\,, 
\end{align}
where $(\Gamma_{\mathbb M})_{AB}\equiv (\Gamma_{\mathbb M})_{AB'}\mathtt C^{B'}{}_B
=(\Gamma_{\mathbb M})_{BA}$. 
In the context of maximal gauged supergravity in $D=6$, 
$\theta^{A'}{}_{\mathbb{M}}$ describes the embedding tensor and 
(\ref{LCmaxSUGRA}) corresponds to the linear constraint, when additional trombone is absent. 

Decomposing $\theta^{A'}{}_{\mathbb M}\to (\theta^{\dot\alpha}{}_M, \theta^{\dot\alpha}{}_\pm, \theta_{\alpha M},\theta_{\alpha \pm})$, 
it is straightforward to show
\begin{align}
\label{}
[D, \theta^{\dot \alpha }{}_M]&=\frac 12 \theta^{\dot \alpha }{}_M \,, \qquad 
[D, \theta^{\dot\alpha }{}_{+}]=\frac 32 \theta^{\dot \alpha }{}_+ \,, \quad \ \
[D, \theta^{\dot \alpha }{}_-]=-\frac 12 \theta^{\dot \alpha }{}_- \,, \notag \\
[D, \theta_{\alpha M}]&=-\frac 12 \theta_{\alpha M} \,, \quad \
[D, \theta_{\alpha +}]=+\frac 12 \theta_{\alpha +} \,, \quad 
[D, \theta_{\alpha -}]=-\frac 32 \theta_{\alpha -} \,.
\end{align}
Note that each of 
$\theta^{\dot \alpha }{}_M$  and $\theta_{\alpha M}$ contains 64 components, 
so that these are reducible.  In the rest of this subsection, we uncover
its irreducible components.

To this aim, we note that the ${\rm SO}(4,4)$ gamma matrices take the following chiral form
\begin{align}
\label{}
(\mathbf \Gamma_M)^{A'}{}_{B'}=(\mathbf \Gamma_M)^{A'B}\delta_{BB'}= \left(
\begin{array}{cc}
      &  (\gamma_M)^{\dot \alpha \beta}  \\
   (\gamma_M)_{\alpha \dot\beta}   &   
\end{array}
\right)\,.  
\end{align}
Spinor indices $\alpha , \dot \alpha$ are raised and lowered by 
$\mathbb C_{\alpha\beta}$, $\mathbb C^{\dot\alpha\dot \beta}$ and their inverse matrices, so that 
$ (\gamma_M)^{\dot \alpha}{}_\beta= (\gamma_M)_\beta{}^{\dot \alpha}=(\gamma_M)^{\dot \alpha \gamma}\mathbb C_{\gamma \beta}$. 
In this representation, 
the condition (\ref{LCmaxSUGRA}) decomposes into 
\begin{align}
\label{}
\sqrt 2 \theta_{\alpha}{}^--(\gamma_M)_{\alpha\dot \beta}\theta^{\dot\beta M}=0\,, \qquad 
\sqrt 2 \theta ^{\dot \alpha +}+(\gamma_M)^{\dot\alpha \beta}\theta_\beta{}^{M}=0\,. 
\end{align}
This suggests us to define
\begin{align}
\label{vartheta}
\vartheta^{\dot \alpha M}\equiv \theta^{\dot \alpha M}-
\frac{\sqrt 2}8(\gamma^M)^{\dot \alpha \beta}\theta_\beta{}^{-}\,,\qquad 
 \vartheta_{\alpha}{}^M\equiv \theta_{\alpha}{}^M+
 \frac{\sqrt 2}{8}(\gamma^M)_{\alpha\dot\beta }\theta^{\dot\beta+}
  \,, 
\end{align}
each of which obeys the following eight restrictions
\begin{align}
\label{LC_N4}
(\gamma_M)_{\alpha\dot\beta}\vartheta^{\dot\beta M}=0 \,, \qquad 
(\gamma_M)^{\dot \alpha \beta}\vartheta_{\beta}{}^{M}=0 \,.
\end{align}
These quantities thus satisfy 
\begin{align}
\label{}
[D, \vartheta_{\alpha M}]&=-\frac 12 \vartheta_{\alpha M} \,, \quad 
[D, \vartheta^{\dot\alpha}{}_{M}]=+\frac 12 \vartheta^{\dot\alpha}{}_{M} \,.
\end{align}
Accordingly, we obtain the following branching rule
\begin{align}
\label{SO55branching_VS}
{\bf 144}_c \to &\ (\vartheta_{\alpha M}\in {\bf 56}_c^{(-1)} )\oplus ({\vartheta^{\dot\alpha}{}_M\in {\bf 56}_s^{(+1)}})\oplus
({\theta_{\alpha+}\in {\bf 8}_s^{(+1)}})  \notag \\
&\ \oplus ({\theta^{\dot\alpha}{}_-\in {\bf 8}_c^{(-1)}}) 
\oplus (\theta^{\dot\alpha}{}_+\in {\bf 8}_c^{(+3)})\oplus(\theta_{\alpha-}\in {\bf 8}_s^{(-3)})\,.
\end{align}

\subsection{Truncation of the quadratic constraints}

Let us next move on to the discussion of the quadratic constraints on the embedding tensor. 
Supersymmetry for the ($2,2$) theory requires the embedding tensor to sit in the ${\bf 144}_c$ 
representations of ${\rm SO}(5,5)$, whose branching under $\mathbb R^+ \times {\rm SO}(4,4)$ was already studied to discuss the implicationes of the linear constraint in (\ref{LCmaxSUGRA}).
Our following task is to obtain the explicit
branching rule for the quadratic constraints on the embedding tensor.

The truncation to ${\cal N}=(1,1)$ theory amounts to projecting out the anti-chiral spinor representations in (\ref{SO55branching_VS})
\begin{align}
\label{}
{\bf 144}_c \to {\bf 56}_c^{(-1)} \oplus \xcancel{{\bf 56}_s^{(+1)}} \oplus \xcancel{{\bf 8}_s^{(+1)}} \oplus {\bf 8}_c^{(-1)} \oplus {\bf 8}_c^{(+3)}
\oplus \xcancel{{\bf 8}_s^{(-3)}}\ .
\end{align}
Accordingly, three different contributions to the embedding tensor survive the truncation, which we conventionally label\footnote{
In the body of the text, these terms possess the ${\rm SO}(4,4)$ fundamental vector indices, instead of chiral spinor ones. 
These indices, however, can be switched at will, thanks to the triality property of ${\rm SO}(4,4)$. 
}
\begin{align}
\label{}
f_{\dot \alpha \dot \beta \dot\gamma} \in {\bf 56}_c^{(-1)}\,, \qquad 
\xi_{\dot \alpha} \in {\bf 8}_c^{(-1)} \,, \qquad 
\zeta_{\dot \alpha } \in {\bf 8}_c^{(+3)} \,. 
\end{align}
$f_{\dot \alpha \dot \beta \dot\gamma}$ represents the gaugings inside ${\rm SO}(4,4)$, whereas 
$\xi_{\dot \alpha}$ is related to the gauging of the scaling symmetry $\mathbb R^+$. 
Finally $\zeta_{\dot \alpha}$ describes instead the massive deformation \cite{Janssen:2001hy,Behrndt:2001ab}. 
We shall illustrate below how these pieces show up from $\theta^{A'}{}_M$. 

Since $\vartheta_{\alpha M}$ introduced in (\ref{vartheta}) belongs to ${\bf 56}_c^{(-1)}$, it can be 
transferred to $f_{\dot\alpha\dot\beta\dot\gamma}$. One can achieve this by the following map 
\begin{align}
\label{ftheta}
f^{\dot\alpha\dot\beta\dot\gamma} \equiv  (\gamma_{MN})^{\dot\alpha\dot\beta}
(\gamma^N)^{\dot\gamma \beta}\vartheta_\beta{}^{M} \ , \qquad 
f^{\dot\alpha\dot\beta\dot\gamma}=f^{[\dot\alpha\dot\beta\dot\gamma]} \ ,
\end{align}
where the second property comes from the linear constraint 
$(\gamma_M)^{\dot \alpha \beta}\vartheta_\beta{}^{M}=0$. 
The relation (\ref{ftheta}) can be inverted to give 
\begin{align}
\label{vartheta_f}
\vartheta_\alpha {}^M=\frac 1{48}f^{\dot\alpha\dot\beta\dot \gamma}(\gamma^{MN})_{\dot\alpha\dot\beta} (\gamma_N)_{\alpha \dot \gamma}
=\frac 1{48}f^{\dot \alpha\dot\beta\dot\gamma}(\gamma_{\dot\alpha\dot\beta\dot\gamma})_\alpha{}^M \ , 
\end{align}
where 
$(\gamma_{\dot\alpha\dot\beta\dot\gamma})_\alpha{}^M\equiv (\gamma^{MN})_{[\dot\alpha\dot\beta} (\gamma_{|N})_{\alpha| \dot \gamma]}$. It is also useful to record
\begin{align}
\label{}
f^{\dot\alpha\dot\beta\dot\gamma}(\gamma^{MN})_{\dot\alpha\dot\beta} =
-16(\gamma^{[M})^{\dot\gamma \beta}\vartheta_\beta{}^{N]} 
=-16(\gamma^{[M}\vartheta^{N]})^{\dot\gamma} \ .
\end{align}
For the rest of the components ${\bf 8}_c^{(-1)}$ and $ {\bf 8}_c^{(+3)}$,  the 
map is simply given by 
\begin{align}
\label{}
\xi^{\dot \alpha} \equiv 2\sqrt 2 \theta^{\dot\alpha +} \ , \qquad 
\zeta ^{\dot \alpha } \equiv \theta ^{\dot \alpha -} \ , 
\end{align}

Quadratic constraints on the embedding tensor in maximal gauged supergravity read \cite{Bergshoeff:2007ef} 
\begin{subequations}
\label{QCN8}
\begin{align}
{\cal Q}_{(1)}^{AB}\equiv \ \ &\eta_{\mathbb{MN}}\theta^{A\mathbb M}\theta^{B\mathbb N}=0\ ,\label{QCN81} \\
{\cal Q}_{(2)}^{\mathbb{MNP}}\equiv\ \ &\theta^{A\mathbb M}\theta^{B[\mathbb N}(\Gamma^{\mathbb P]})_{AB}=0\label{QCN82} \ .   
\end{align}
\end{subequations}
The first quadratic constraints (\ref{QCN81}) split into 
${\cal Q}_{(1)}^{\dot \alpha \dot \beta}=0$ and 
${\cal Q}_{(1)\alpha\beta}=0$. The former reduces to 
\begin{align}
\label{}
{\cal Q}_{(1)}^{\dot \alpha \dot \beta}=
{\frac 1{\sqrt 2}}\xi^{(\dot \alpha}\zeta ^{\dot \beta)} \ . 
\end{align} 
This recovers $\mathbf{35}_{c}^{(+2)}\oplus\mathbf{1}^{(-2)}$ in (\ref{eq:QC}). 
Contraction of $\mathbb C^{\alpha \beta} $ to ${\cal Q}_{(1)\alpha\beta}$ yields
\begin{align}
\label{QCtruncation1:tr}
\mathbb C^{\alpha \beta} {\cal Q}_{(1)\alpha\beta}=
\frac 1{48}\left(f^{\dot\alpha\dot \beta\dot \gamma}f_{\dot\alpha\dot \beta\dot \gamma}+\frac 32 \xi_{\dot \alpha}\xi^{\dot \alpha}\right) \ . 
\end{align}
The trace-free part in ($\alpha, \beta$) for ${\cal Q}_{(1)\alpha\beta}$   can be computed by the contraction to  
$(\gamma_{MNPQ})^{\alpha \beta}$,  which now becomes 
\begin{align}
\label{}
(\gamma_{MNPQ})^{\alpha \beta} {\cal Q}_{(1)\alpha\beta}&\, =
-\frac 1{256}f^{\dot \mu \dot \alpha \dot \beta}f_{\dot \mu}{}^{\dot
\gamma \dot\delta}(\gamma_{MNPQ})^{\alpha \beta} (\gamma_{\dot \alpha\dot\beta\dot\gamma\dot\delta})_{\alpha \beta} 
\notag \\ 
&\,=-\frac 3{32} \left(f^{\dot \mu[ \dot \alpha\dot\beta}f^{\dot\gamma\dot\delta]}{}_{\dot\mu}+
\frac 23 \xi^{[\dot \alpha}f^{\dot\beta\dot \gamma\dot\delta]} \right)
(\gamma_{[\dot\alpha\dot\beta})_{[MN}(\gamma_{\dot\gamma\dot\delta]})_{PQ]} \ , 
\end{align}
where we have used 
$(\gamma_{MNPQ})^{\alpha \beta} (\gamma_{\dot \alpha\dot\beta\dot\gamma\dot\delta})_{\alpha \beta} =24(\gamma_{[\dot\alpha\dot\beta})_{[MN}(\gamma_{\dot\gamma\dot\delta]})_{PQ]}$. 
Noting that $(\gamma_{[\dot\alpha\dot\beta})_{[MN}(\gamma_{\dot\gamma\dot\delta]})_{PQ]}$
is anti-selfdual in indices $\dot\alpha\dot\beta\dot\gamma\dot\delta$, we obtain 
\begin{align}
\label{}
\left.f_{\dot \mu [\dot \alpha \dot \beta}f_{\dot \gamma \dot \delta ]}{}^{\dot \mu}
-\frac 23 f_{[\dot \alpha\dot\beta\dot\gamma}\xi_{\dot \delta]} 
\right|_{\rm ASD} =0 \,,
\end{align}
This is nothing but the piece in the {${\bf 35}^{(-2)}_{v}$} of the QC (\ref{eq:QC}).

The second set of the quadratic constraint ${\cal Q}_{(2)}^{\mathbb{MNP}}=0$ decomposes into
various components. A quick computation shows
\begin{align}
\label{}
{\cal Q}_{(2)}^{- - +}=&\,{\frac 1{2}} \xi^{\dot \alpha}\zeta _{\dot \alpha}=0\ , \\
{\cal Q}_{(2)}^{+ - +}=&\, \frac 1{4\sqrt 2}\xi^{\dot \alpha}\xi _{\dot \alpha}=0\ , 
\end{align}
yielding ${\bf 1}^{(+2)}$ and ${\bf 1}^{(-2)}$ in (\ref{eq:QC}). 
Combined with (\ref{QCtruncation1:tr}), one gets a condition inherent to 
the truncation of maximal gauged supergravity [the first of (\ref{eq:QC-extra})]
\begin{align}
\label{}
f^{\dot\alpha\dot \beta\dot \gamma}f_{\dot\alpha\dot \beta\dot \gamma}=0 \ . 
\end{align}
Similarly, we have 
\begin{align}
\label{}
{\cal Q}_{(2)}^{-MN}=-
{\frac 1{8}}\left(f_{\dot \alpha\dot \beta\dot \gamma}\zeta^{\dot \alpha }
-\xi_{[\dot \beta}\zeta _{\dot \gamma]}\right)(\gamma^{MN})^{\dot \beta\dot \gamma}\ , 
\end{align}
i.e., $f_{\dot \alpha\dot \beta\dot \gamma}\zeta^{\dot \alpha -}-\xi_{[\dot \beta}\zeta _{\dot \gamma]}=0$, 
which corresponds to ${\bf 28}^{(+2)}$ in (\ref{eq:QC}). 
Further computations give rise to 
\begin{align}
\label{}
{\cal Q}_{(2)}^{(MN)-}={\frac 1{48}}
f^{\dot\alpha\dot\beta\dot\gamma }\zeta^{\dot\delta}
(\gamma_{\dot\alpha\dot\beta\dot\gamma\dot\delta})^{MN} \ . 
\end{align}
Since  $(\gamma_{\dot\alpha\dot\beta\dot\gamma\dot\delta})^{MN}$ is self-dual 
in indices $[\dot\alpha\dot\beta\dot\gamma\dot\delta]$,  
we obtain 
\begin{align}
\label{}
\left.f_{[\dot\alpha\dot \beta\dot\gamma}\zeta_{\dot \delta]} \right|_{\rm SD} =0 \ .
\end{align}
This quadratic constraint is also intrinsic to the truncation of maximal gauged supergravity 
[see (\ref{eq:QC-extra})]. 
We obtain also 
\begin{align}
\label{}
{\cal Q}_{(2)}^{[MN]+}(\gamma_{\dot \alpha \dot \beta})_{MN} =\frac 1{2\sqrt 2}
f_{\dot \alpha\dot\beta\dot\gamma}\xi^{\dot \gamma}  \ , 
\end{align}
yielding ${\bf 28}^{(-2)}$ in (\ref{eq:QC}). 
Lastly, one finds 
\begin{align}
\label{}
{\cal Q}_{(2)}^{(MN)+}-\frac 18 \eta^{MN}\eta_{PQ}{\cal Q}_{(2)}^{(PQ)+}
=-\frac{\sqrt 2}{256}\left(f_{\dot \mu [\dot \alpha \dot \beta}f_{\dot \gamma \dot \delta ]}{}^{\dot \mu}
-\frac 23 f_{[\dot \alpha\dot\beta\dot\gamma}\xi_{\dot \delta]} \right)(\gamma^{\dot \alpha\dot\beta\dot\gamma\dot \delta})^{MN} 
\,. 
\end{align}
Using the self-duality of $(\gamma^{\dot \alpha\dot\beta\dot\gamma\dot \delta})^{MN} $, one ends up with 
the {${\bf 35}^{(-2)}_{s}$} in (\ref{eq:QC}), 
\begin{align}
\label{}
\left.f_{\dot \mu [\dot \alpha \dot \beta}f_{\dot \gamma \dot \delta ]}{}^{\dot \mu}
-\frac 23 f_{[\dot \alpha\dot\beta\dot\gamma}\xi_{\dot \delta]} 
\right|_{\rm SD} =0 \ .
\end{align}

This completes the truncation of the quadratic constraints. 
Equations obtained here match exactly the quadratic constraints in ${\cal N}=(1,1)$ theory 
derived by the requirement of consistent gaugings and massive deformations.


\section{Global aspects of the vacua}
\label{app:global}

Most of our string/M-theory vacua turn out to correspond to reductions over a 4D(5D) group manifold (\emph{a.k.a.} a twisted torus).
In this appendix we discuss some details concerning the global issues that may obstruct viewing the aforementioned manifolds as genuinely compact. 
These will determine the global aspects of the flux compactifications yielding the corresponding vacuum solution.

From the consistency of the dimensional reduction, we must have a set of $n$ globally defined \emph{left-invariant} 1-forms $\sigma^a$ ($a=1,\ldots,n$) on the $n$-dimensional internal manifold that satisfy 
\begin{align}
\D \sigma^a=-\frac12 \omega_{bc}{}^a \ \sigma^b\wedge\sigma^c
\ ,
\end{align}
where $\omega_{bc}{}^{a}$ is the metric flux, which will turn out to encode the structure constants of the underlying group structure. 
This condition, together with the constancy of the flux, shows that the internal space is a group manifold, where the components $\omega_{bc}{}^a$ are actually the structure constants of the group, and hence satisfy 
\begin{align}
\label{}
\omega_{ab}{}^c=\omega_{[ab]}{}^c \ , \qquad 
\omega_{[ab}{}^d\omega_{c]d}{}^e =0 \ .  
\end{align}
A necessary condition for the compactification of the group manifold $G$ is that the 
group be \emph{unimodular}, \emph{i.e.} 
\begin{align}
\label{unimodular}
\omega_{ab}{}^b =0 \ . 
\end{align}
The unimodular group $G$ may then admit a discrete and freely acting subgroup $\Gamma$ (\emph{i.e.} free of fixed points), 
permitting  $G/\Gamma$ to be compact. If the unimodularity condition has been dropped, 
the volume of the internal manifold would vary, prohibiting the compactification. 
See e.g., \cite{Grana:2006kf} for a comprehensive analysis on this point. 

In the following subsections we are going to obtain the explicit Maurer-Cartan 1-forms and their global identifications for the various vacua solutions that have been found above in the main text. We shall overall denote the 4D internal type II coordinates by $y^m\equiv\left(\tau,\,x,\,y,\,z\right)$, while the 5D ones in M-theory will be $y^{\hat m}\equiv\left(\tau,\,x,\,y,\,z,\,w\right)$.

\subsubsection*{Massive Type IIA with O6/D6}
For solution 1 in Table \ref{tab:O6-solutions}, the Maurer-Cartan forms are given by
\begin{align}
\sigma^1 
=&\
\D \tau-\beta_3x \D y -\beta_1 y \D z -\beta_2 z \D x
\ ,
&
\sigma^3
=&\
\D y
\ ,
\\
\sigma^2
=&\
\D x
\ ,
&
\sigma^4
=&\
\D z
\ .
\end{align}
Then we can do a compactification with the following identifications:
\begin{align}
\begin{array}{lcccl}
(\tau,x,y,z)
&\simeq&
(\tau+L_0,\ x,\ y,\ z)
&\simeq&
(\tau+\beta_3 L_1 y,\ x+L_1,\ y,\ z)
\\
&\simeq&
(\tau+\beta_1 L_2 z,\ x,\ y+L_2,\ z)
&\simeq&
(\tau+\beta_3 z,\ x,\ y,\ z+L_3)
\ ,
\end{array}
\end{align}
where $L_0,\ldots,L_3$ are some arbitrary real constants.

For solution 2 in Table \ref{tab:O6-solutions}, one can always achieve $\beta_2=\beta_3=0$ 
by a suitable ${\rm O}(3)$ rotation, for which the Maurer-Cartan forms are given by
\begin{align}
\sigma^1 
=&\
\D \tau
\ ,
&
\sigma^3
=&\
\cos(\beta_1\tau)\D y-\sin(\beta_1\tau)\D x
\ ,
\\
\sigma^2
=&\
{\cos(\beta_1\tau)\D x+\sin(\beta_1\tau)\D y}
\ ,
&
\sigma^4
=&\
\D z
\ .
\end{align}
This is just flat space, so we can do a compactification with the following identifications:
\begin{align}
\begin{array}{lcccl}
(\tau,\ x,\ y,\ z)
&\simeq&
(\tau+\tfrac{2\pi}{\beta_1},\ x,\ y,\ z)
&\simeq&
(\tau,\ x+L_1,\ y,\ z)
\\
&\simeq&
(\tau,\ x,\ y+L_2,\ z)
&\simeq&
(\tau,\ x,\ y,\ z+L_3)
\ ,
\end{array}
\label{eq:O6-global}
\end{align}
where $L_0,\ldots,L_3$ are some arbitrary real constants.

For the 3rd solution of Table \ref{tab:O6-solutions}, one can set $\beta=0$ 
by an ${\rm O}(3)$ rotation, for which the Maurer-Cartan forms are given by
\begin{align}
\sigma^1 
=&\
\cos(\alpha x)\D \tau-\sin(\alpha x)\D y
\ ,
&
\sigma^3
=&\
{\sin(\alpha x)\D \tau+\cos(\alpha x)\D y}
\ ,
\\
\sigma^2
=&\
\D x
\ ,
&
\sigma^4
=&\
\D z
\ .
\end{align}
In this case, we also have flat space and the identifications are exactly the same as in the 2nd case, which are given by \eqref{eq:O6-global} 
with the replacement $\beta_1 \to \alpha$.

\subsubsection*{Type IIB with O7/D7}
For solution 1 in Table \ref{tab:O7-solutions}, we find
\begin{align}
\sigma^1 
=&\
\D \tau+\alpha y \D z
\ ,
&
\sigma^3
=&\
\cos(\beta x)\D y-\sin(\beta x)\D z
\ ,
\\
\sigma^2
=&\
\D x
\ ,
&
\sigma^4
=&\
{\cos(\beta x)\D z+\sin(\beta x)\D y}
\ .
\end{align}
Then we can do a compactification with the following identifications:
\begin{align}
\begin{array}{lcccl}
(\tau,\ x,\ y,\ z)
&\simeq&
(\tau+L_0,\ x,\ y,\ z)
&\simeq&
(\tau,\ x+\tfrac{2\pi}{\beta},\ y,\ z)
\\
&\simeq&
(\tau-\alpha L_2 z,\ x,\ y+L_2,\ z)
&\simeq&
(\tau,\ x,\ y,\ z+L_3)
\ ,
\end{array}
\end{align}
where $L_0,\ldots,L_3$ are some arbitrary real constants.

For solution 2 of Table \ref{tab:O7-solutions}, we find
\begin{align}
\sigma^1 
=&\
\frac{1}{\sqrt{\alpha^2+\beta^2}}\left(
	\beta \D \tau
	+\alpha\cos(\sqrt{\alpha^2+\beta^2}\ y )\D x
	-\alpha\sin(\sqrt{\alpha^2+\beta^2}\ y )\D z
	\right)
\ ,
\\
\sigma^2
=&\
\frac{1}{\sqrt{\alpha^2+\beta^2}}\left(
	-\alpha \D \tau
	+\beta\cos(\sqrt{\alpha^2+\beta^2}\ y )\D x
	-\beta\sin(\sqrt{\alpha^2+\beta^2}\ y )\D z
	\right)
\ ,
\\
\sigma^3
=&\
\D y
\ ,
\qquad
\sigma^4
=
{\cos(\sqrt{\alpha^2+\beta^2}y)\D z+\sin(\sqrt{\alpha^2+\beta^2} y)\D x}
\ .
\end{align}
The manifold may then be made compact by performing the following identifications
\begin{align}
\begin{array}{lcccl}
(\tau , x, y , z) &\simeq& (\tau +L_0, x, y , z) &\simeq &
(\tau , x+L_1, y , z) \notag \\ &\simeq &
(\tau , x, y+2\pi/\sqrt{\alpha^2+\beta^2}, z) &\simeq &
(\tau , x, y , z+L_3)\,.  
\end{array}
\end{align}

For solution 3 of Table \ref{tab:O7-solutions}, we find
\begin{align}
\label{}
\sigma^1=&\D \tau \,, &
\sigma^3=& \cos(\alpha \tau+\beta x)\D y-\sin(\alpha \tau+\beta x)\D z  \,, \\
\sigma^2=&\D x \,, &
\sigma^4=&\cos(\alpha \tau+\beta x)\D z+\sin (\alpha \tau+\beta x)\D y  \,. 
\end{align}
Then, defining
\begin{align}
u\equiv \frac{\alpha\tau +\beta x}{\sqrt{\alpha^2+\beta^2}}
\ ,
\qquad
v\equiv \frac{-\beta\tau +\alpha x}{\sqrt{\alpha^2+\beta^2}}
\ ,
\end{align}
we can do a compactification with the following identifications:
\begin{align}
\begin{array}{lcccl}
(u,v,y,z)
&\simeq&
\left(u+\tfrac{2\pi}{\sqrt{\alpha^2+\beta^2}},v,y,z\right)
&\simeq&
(u,v+L_1,y,z)
\\
&\simeq&
(\tau,x,y+L_2,z)
&\simeq&
(\tau,x,y,z+L_3)
\ ,
\end{array}
\end{align}
where $L_0,\ldots,L_3$ are some arbitrary real constants.

\subsubsection*{Massive Type IIA with O8/D8}

For solution 1 in Table \ref{tab:O8-solutions}, we find
\begin{align}
\sigma^1 
=&\
\D \tau
\ ,
&
\sigma^3
=&\
{\cos(\alpha z)\D y-\sin(\alpha z)\D x}
\ ,
\\
\sigma^2
=&\
\cos(\alpha z)\D x+\sin(\alpha z)\D y
\ ,
&
\sigma^4
=&\
\D z
\ .
\end{align}
Then we can do a compactification with the following identifications:
\begin{align}
\begin{array}{lcccl}
(\tau,x,y,z)
&\simeq&
(\tau+L_0,x,y,z)
&\simeq&
(\tau,x+L_1,y,z)
\\
&\simeq&
(\tau,x,y+L_2,z)
&\simeq&
(\tau,x,y,z+\frac{2\pi}{\alpha})
\ .
\end{array}
\end{align}

\subsubsection*{Type IIB with O9/D9}

For solution 1 in Table \ref{tab:O9-solutions}, we find
\begin{align}
\sigma^1 
=&\
\frac{1}{\sqrt{\alpha^2+\beta^2}}\left(
	\beta \D \tau
	+\alpha\cos(\sqrt{\alpha^2+\beta^2}\ y ){\D x}
	-\alpha\sin(\sqrt{\alpha^2+\beta^2}\ y )\D z
	\right)
\ ,
\\
\sigma^2
=&\
\frac{1}{\sqrt{\alpha^2+\beta^2}}\left(
	-\alpha \D \tau
	+\beta\cos(\sqrt{\alpha^2+\beta^2}\ y )\D x
	-\beta\sin(\sqrt{\alpha^2+\beta^2}\ y )\D z
	\right)
\ ,
\\
\sigma^3
=&\
\D y
\ ,
\qquad 
\sigma^4
=
\cos(\sqrt{\alpha^2+\beta^2}\ y )\D z
+\sin(\sqrt{\alpha^2+\beta^2}\ y )\D x
\ .
\end{align}
Then we can do a compactification with the following identifications:
\begin{align}
\begin{array}{lcccl}
(\tau,x,y,z)
&\simeq&
(\tau+L_0,x,y,z)
&\simeq&
(\tau,x+L_1,y,z)
\\
&\simeq&
(\tau,x,y+\frac{2\pi}{\sqrt{\alpha^2+\beta^2}},z)
&\simeq&
(\tau,x,y,z+L_3)
\ .
\end{array}
\end{align}

For solution 2 of Table \ref{tab:O9-solutions}, we find
\begin{align}
\sigma^1 
=&\
\cos(\sqrt{\beta(\alpha+\beta)}\ z )\D \tau
-\sin(\sqrt{\beta(\alpha+\beta)}\ z )\D y
\ ,
\qquad
\sigma^2
= 
\D x
\ ,
\\
\sigma^3
=&\
\sqrt{\frac{\beta}{\alpha+\beta}}\left(
	\cos(\sqrt{\beta(\alpha+\beta)}\ z )\D y
	+\sin(\sqrt{\beta(\alpha+\beta)}\ z )\D \tau
	\right)
\ ,
\\
\sigma^4
=&\
\D z
+
\sqrt{\frac{-\alpha}{\alpha+\beta}}\left(
	\cos(\sqrt{\beta(\alpha+\beta)}\ z )\D \tau
	-\sin(\sqrt{\beta(\alpha+\beta)}\ z )\D y
	\right)
\ .
\end{align}
Then we can do a compactification with the following identifications:
\begin{align}
\begin{array}{lcccl}
(\tau,x,y,z)
&\simeq&
(\tau+L_0,x,y,z)
&\simeq&
(\tau,x+L_1,y,z)
\\
&\simeq&
(\tau,x,y+L_2,z)
&\simeq&
(\tau,x,y,z+\frac{2\pi}{\sqrt{\beta(\alpha+\beta)}})
\ .
\end{array}
\end{align}

For solution 3 of Table \ref{tab:O9-solutions}, we find
\begin{align}
\sigma^1 
=&\
\D \tau
\ ,
&
\sigma^3
=&\
\cos(\alpha\tau+\beta x)\D y-\sin(\alpha\tau+\beta x)\D z
\ ,
\\
\sigma^2
=&\
\D x
\ ,
&
\sigma^4
=&\
{\cos(\alpha\tau+\beta x)\D z+\sin(\alpha\tau+\beta x)\D y}
\ .
\end{align}
This solution is the same as solution 3 of type IIB with O7/D7 in Table \ref{tab:O7-solutions}, and the global identifications are also equal.

\subsubsection*{M-theory with KKO6/KK6}

For solution 1 in Table \ref{tab:KKO6-solutions}, we find
\begin{align}
\label{}
\sigma^1 =&\cos (\alpha x/\beta)\D \tau -\sin (\alpha x/\beta)\D y \,, \qquad 
\sigma^2=\frac{\D x-\beta \cos (\alpha x/\beta)\D y-\beta \sin (\alpha x/\beta)\D \tau }{\sqrt{\alpha^2+\beta^2}}\,, \notag \\
\sigma^3=&\frac{\alpha \cos  (\alpha x/\beta)\D u+\alpha \sin (\alpha x/\beta)\D \tau }{\sqrt{\alpha^2+\beta^2}}\,, \qquad 
\sigma^4=\D z \,, \qquad \sigma^5=\D w \,. 
\end{align}
We can do a compactification with the following identifications:
\begin{align}
\label{}
(\tau , x, y , z,w) &\simeq (\tau +L_0, x, y , z,w) \simeq 
(\tau , x+2\pi \beta/\alpha, y , z,w) \notag \\
&\simeq 
(\tau , x, y+L_2,  z,w) \simeq 
(\tau , x, y , z+L_3,w)\simeq (\tau, x, y, z, w+L_4)\ .  
\end{align}

For solution 3 in Table \ref{tab:KKO6-solutions}, we have 
\begin{align}
\label{}
\sigma^1 =&\D \tau \,, \qquad \sigma^2 =\cos (\alpha \tau)\D x+\sin(\alpha \tau)\D y \,, \qquad 
\sigma^3=\cos (\alpha \tau)\D y-\sin(\alpha \tau)\D x \,, \notag \\
\sigma^4=&\cos(\beta \tau) \D z+\sin(\beta \tau)\D w \,, \qquad 
\sigma^5=\cos(\beta \tau)\D w-\sin(\beta \tau)\D w \,. 
\end{align}
One can assume $|\beta|\ge |\alpha|$ without loss of generality. Then, 
the global compactification is possible iff $\beta= n \alpha $ ($n \in \mathbb Z$) with 
\begin{align}
\label{}
(\tau , x, y , z,w) &\simeq (\tau +2\pi/\alpha, x, y , z,w) \simeq 
(\tau , x+L_1, y , z,w) \notag \\
&\simeq 
(\tau , x, y+L_2,  z,w) \simeq 
(\tau , x, y , z+L_3,w)\simeq (\tau, x, y, z, w+L_4)\,.  
\end{align}

\small


\providecommand{\href}[2]{#2}\begingroup\raggedright\endgroup

\end{document}